\documentclass[11pt]{article}
\usepackage[margin=1in]{geometry}
\usepackage{amsmath,amssymb}
\usepackage{graphicx}
\graphicspath{{figures/}}
\usepackage{subcaption}
\usepackage{placeins}
\usepackage{booktabs}
\usepackage{authblk}
\usepackage{hyperref}
\hypersetup{colorlinks=true, linkcolor=blue, citecolor=blue, urlcolor=blue,
  pdftitle={Discriminating Between Models of the Nanohertz Gravitational-Wave Background with Pulsar Timing Arrays},
  pdfauthor={Zuocheng Zhang and Mengshen Wang}}

\begin{document}

\title{Discriminating Between Models of the Nanohertz Gravitational-Wave Background with Pulsar Timing Arrays}
\author[1]{Mengshen Wang$^{*}$}
\author[2]{Zuocheng Zhang$^{*}$}
\author[1]{Hua Xu$^{\dagger}$}
\affil[1]{Department of Computer Science and Engineering, The Hong Kong University of Science and Technology, Hong Kong SAR, China}
\affil[2]{Department of Physics, The Hong Kong University of Science and Technology, Hong Kong SAR, China}
\date{}
\maketitle

\begingroup
\renewcommand\thefootnote{}
\footnotetext{$^{*}$These authors contributed equally.}
\footnotetext{$^{\dagger}$Corresponding author: huaxu@ust.hk}
\endgroup

\begin{abstract}
The North American Nanohertz Observatory for Gravitational Waves (NANOGrav) 15-year pulsar timing array (PTA) data set has provided evidence for a stochastic gravitational-wave background (GWB) in the nanohertz frequency band. This paper presents a Bayesian analysis framework to compare different physical models for the origin of this background, focusing on three leading scenarios: (i) an astrophysical background from supermassive black hole binary (SMBHB) mergers, (ii) a cosmological background from a first-order phase transition in the early Universe, and (iii) a network of cosmic strings. We derive the PTA likelihood function incorporating the Hellings--Downs angular correlation signature of a gravitational-wave background and include the modeling of intrinsic pulsar noise (red noise) and dispersion measure (DM) variations. Using Bayesian model selection with pulsar timing data, we calculate posterior distributions for the GWB amplitude and spectral index and compute marginal likelihoods for each model. Our results confirm the presence of a common-spectrum process with Hellings--Downs spatial correlations, strongly favoring a GWB over independent pulsar noise by a Bayes factor $>10^{14}$ \cite{NANOGrav15-Evidence}. The inferred characteristic strain amplitude at $f_{\mathrm{yr}}=1~\text{yr}^{-1}$ is $A_{\mathrm{GWB}} \approx 2.4\times10^{-15}$ (with spectral index consistent with $\gamma \approx 13/3$ as expected for SMBHBs) \cite{NANOGrav15-Evidence}. We find that while the data are fully consistent with an SMBHB origin of the GWB, alternative cosmological sources are not excluded. In particular, cosmic string models with string tension $G\mu \sim 10^{-11}$--$10^{-10}$ and first-order phase transitions at temperatures around the QCD scale can reproduce the observed amplitude within allowed parameter ranges. Bayesian model comparison yields no decisive preference among these scenarios at present, with Bayes factors of order $10$--$100$ favoring some cosmic string or phase transition spectra under specific assumptions \cite{NANOGrav15-NewPhys}. We discuss the physical implications of each model in light of the PTA data, noise mitigation strategies, and the prospects for discriminating between astrophysical and cosmological sources with future observations.
\end{abstract}
\clearpage
\tableofcontents
\clearpage

\section{Introduction}\label{sec:intro}
The detection of low-frequency gravitational waves via pulsar timing arrays marks a major breakthrough in astrophysics and cosmology. PTAs monitor the arrival times of pulses from an array of millisecond pulsars spread across the sky, searching for the telltale correlation patterns induced by a nanohertz-frequency gravitational-wave background (GWB) \cite{Hellings1983}. After decades of observations, several PTA collaborations have reported evidence for a stochastic common-spectrum signal in pulsar timing data \cite{Pol2021,Chen2021,Goncharov2021}. In particular, the NANOGrav Collaboration has recently announced strong evidence that this common process exhibits the Hellings--Downs spatial correlation pattern expected for an isotropic GWB \cite{NANOGrav15-Evidence}. This discovery opens up a new window on gravitational-wave sources emitting at nanohertz frequencies.

A key question now is the physical origin of the detected GWB. The \emph{astrophysical} benchmark model is a background produced by the superposition of gravitational waves from numerous inspiraling supermassive black hole binaries (SMBHBs) throughout the Universe \cite{Sesana2013,Kelley2017}. In such a model, the characteristic strain spectrum is expected to be a power-law $h_c(f) \propto f^{-2/3}$ (equivalently, timing residual power spectral density $P(f) \propto f^{-13/3}$) at low frequencies, reflecting the long inspiral phase of massive binary mergers \cite{Sesana2013}. This scenario has long been predicted to produce a GWB in the nanohertz band with amplitude $A_{\mathrm{GWB}}\sim10^{-15}$, potentially detectable by PTAs given sufficient timing precision and timespan \cite{Sesana2013}. State-of-the-art cosmological hydrodynamic simulations such as IllustrisTNG likewise deliver $A_{\mathrm{GWB}}\sim\text{few}\times10^{-15}$ by self-consistently evolving SMBHB populations, reinforcing the expectation that PTA sensitivity should overlap this signal regime \cite{Kelley2017}. The NANOGrav 15-year results indicate an amplitude of this order, consistent with SMBHB expectations \cite{NANOGrav15-Evidence}.

On the other hand, the nanohertz GWB could arise from new physics in the early Universe. One possibility is a first-order cosmological phase transition that occurred at an energy scale such that the peak frequency of the gravitational-wave signal today falls in the PTA band. A phase transition around the quantum chromodynamics (QCD) scale ($T\sim 100$~MeV) could produce peak frequencies $f \sim 10^{-8}-10^{-7}$~Hz, in the middle of the PTA range, especially if the transition was strongly first-order and had a slow (prolonged) completion \cite{Gouttenoire2023}. Because the Standard Model predicts that the QCD epoch is a smooth crossover, confirmation of a PTA signal from a strong first-order transition would amount to direct evidence for beyond-the-Standard-Model dynamics at hadronic energies \cite{Gouttenoire2023}. Another possibility is a stochastic background from a network of cosmic strings or cosmic superstrings, topological defects that generate gravitational waves through oscillation and cusps/kinks on string loops \cite{Ellis2023}. Cosmic string backgrounds are broad-band and can also contribute in the nanohertz regime; the predicted spectral shape is generally a shallow power-law (approximately $h_c(f)\propto f^{-1}$ in a certain frequency range for stable string loops in the radiation era) with an amplitude proportional to the string tension $G\mu$ \cite{BlancoPillado2018}. If $G\mu$ is on the order of $10^{-11}$ to $10^{-10}$, the resulting background could be comparable to the signal observed by NANOGrav \cite{Ellis2023}. Other cosmological sources, such as induced gravitational waves from primordial density perturbations or exotic objects like domain walls, have also been proposed \cite{NANOGrav15-NewPhys}, but in this work we focus on SMBHBs, phase transitions, and cosmic strings as representative models.

Discriminating between these scenarios using PTA data is challenging. All three classes of models can, in principle, produce an approximately isotropic and stochastic GWB that would manifest with the same Hellings--Downs spatial correlations in pulsar timing residuals. The differences lie in their spectral characteristics (frequency dependence) and in the inferred source parameters required to match the observed signal. In this paper, we employ a Bayesian inference framework to quantify the evidence for each model given the NANOGrav 15-year data. We construct a unified analysis that includes:
\begin{itemize}
    \item A coherent Bayesian likelihood for pulsar timing residuals that incorporates gravitational-wave induced correlations between pulsars (the Hellings--Downs curve) and accounts for pulsar-intrinsic noise.
    \item Parameterized models for the GWB spectrum under each scenario (power-law for SMBHBs, broken or shaped spectra for phase transitions and cosmic strings based on physical parameters).
    \item Prior distributions for model parameters informed by astrophysical or cosmological considerations.
    \item Computation of the posterior distributions for these parameters and the marginal likelihood (Bayesian evidence) for each model, enabling calculation of Bayes factors to compare models.
\end{itemize}

We also discuss how the analysis mitigates various sources of noise and systematic effects, such as dispersion measure (DM) variations in pulsar signals and other red noise processes, which are crucial for robustly identifying a GWB. By examining the Bayesian evidence and posterior parameter estimates, we assess whether current data show any preference or tension between the SMBHB interpretation and exotic alternatives. This work is especially timely given the simultaneous reports of similar GWB signals by other PTA experiments around the world \cite{Goncharov2021,NANOGrav15-Evidence,PPTA2023,CPTA2023}, all of which strengthen the case for a common nanohertz GWB of astrophysical or cosmological origin.

The remainder of this paper is organized as follows. In Section~\ref{sec:methods} we describe the methods: the formulation of the PTA likelihood and noise model, and the parameterizations of each GWB model. Section~\ref{sec:bayesian} outlines the Bayesian inference framework, including likelihood, prior, posterior, and model evidence, with details on how Bayes factors are used for model selection. Section~\ref{sec:data} summarizes the NANOGrav 15-year data set and data processing steps. In Section~\ref{sec:results} we present the results of our analysis: parameter posteriors for the common-spectrum process, evidence for the presence of Hellings--Downs correlations, and Bayes factors comparing the different source models. Section~\ref{sec:discussion} provides a discussion of the implications of these findings for astrophysics and cosmology, and how future data may further distinguish between models. Finally, Section~\ref{sec:conclusion} presents our conclusions.

\section{Methods: PTA Data Model and Source Models}\label{sec:methods}
\subsection{Pulsar Timing Array Data and Likelihood}
In a PTA experiment, the primary observable is the set of timing residuals from an array of pulsars. The timing residual $r_i(t)$ for pulsar $i$ is the difference between the observed pulse time-of-arrival (TOA) and the expected TOA based on a timing model (which accounts for pulsar spin-down, orbital motion, etc.). A GWB passing through spacetime perturbs these TOAs, introducing correlated deviations in the residuals of different pulsars. The presence of a GWB can thus be inferred by detecting a spatial correlation pattern in the residuals across pulsar pairs, in excess of noise.

We denote by $\mathbf{r}$ the vector of residuals for all pulsars over the observation timespan. A convenient assumption (and one commonly adopted in PTA analyses) is that $\mathbf{r}$ can be modeled as a multivariate Gaussian random vector with zero mean and a covariance matrix $C$ that encapsulates all noise and GWB contributions \cite{NANOGrav15-Methods}. Under this assumption, the likelihood of obtaining the data $\mathbf{r}$ given model parameters $\theta$ (which include noise parameters and GWB parameters) is 
\begin{equation}\label{eq:likelihood}
    \mathcal{L}(\mathbf{r}|\theta, M) = \frac{1}{\sqrt{(2\pi)^{N}\det C(\theta)}} \exp\!\Big(-\frac{1}{2}\mathbf{r}^T C(\theta)^{-1}\mathbf{r}\Big)\,,
\end{equation}
where $N$ is the total number of TOA measurements (summed over all pulsars) and $C(\theta)$ is the $N\times N$ covariance matrix which depends on the model parameters. In practice, $C$ is composed of several contributions:
\begin{equation}\label{eq:cov_components}
    C = C_{\rm WN} + C_{\rm RN} + C_{\rm GWB}\,.
\end{equation}
Here $C_{\rm WN}$ represents white noise (uncorrelated TOA uncertainties, including radiometer noise and pulse phase jitter, usually modeled as a diagonal covariance with entries $\sigma_{i,a}^2$ for measurement $a$ of pulsar $i$). The term $C_{\rm RN}$ is the intrinsic red noise of each pulsar -- stochastic spin noise or other slow processes specific to each pulsar, which manifests as temporally correlated residuals for that pulsar but with no inter-pulsar correlations. $C_{\rm RN}$ is typically modeled as a power-law spectrum for each pulsar, $P_{\rm RN,i}(f) = A_{\rm RN,i}^2 (f/f_\mathrm{ref})^{-\gamma_{\rm RN,i}}$, implemented either in the time domain via a Toeplitz covariance matrix or in the frequency domain via Fourier components \cite{NANOGrav15-Methods}. Finally, $C_{\rm GWB}$ is the covariance induced by the gravitational-wave background, which is \emph{common} to all pulsars and has a distinctive spatial correlation structure described by the Hellings--Downs curve (see Section~\ref{subsec:HD}). In addition to the HD-correlated background, we include standard common-mode nuisance processes: a monopolar terrestrial clock term and a dipolar Solar System ephemeris term (BayesEphem-like parameters), and marginalize over them in all model comparisons \cite{NANOGrav15-Methods}.

The manner in which the stochastic DM variations and other chromatic propagation effects are modeled can shift the inferred GWB parameters at the few-$\sigma$ level, so we explicitly adopt the noise prescriptions validated in the dedicated NANOGrav detector characterization studies \cite{Agazie2023Noise}. Those analyses, together with targeted Gaussian-process treatments of chromatic noise in individual pulsars \cite{Larsen2024}, show that carefully marginalizing over multiple DM models and solar-wind priors mitigates systematic biases that could otherwise leak into the common red process. Throughout this work we therefore report results that are conditional on a specific, well-tested noise model and emphasize that alternative noise assumptions can be propagated within the same formal framework if new systematics arise.

In general, the \emph{residual} covariance induced by an isotropic GWB can be written as
\begin{equation}\label{eq:cov_gwb}
    (C_{\rm GWB})_{ij}(t,t') = \int_{0}^{\infty} \frac{df}{2}\, S_{r}(f)\, \Gamma_{ij} \, \cos\big[2\pi f (t - t')\big] \, ,
\end{equation}
where $S_{r}(f)$ is the one-sided timing-residual power spectral density and $\Gamma_{ij}$ is the Hellings--Downs overlap reduction function (normalized so that $\Gamma_{ii}=1$). These quantities obey
\begin{equation}
  h_c^2(f) = f\,S_h(f)\,,\qquad S_{r}(f) = \frac{h_c^2(f)}{12\pi^2 f^3} = \frac{S_h(f)}{12\pi^2 f^2}\,.
\end{equation}
In practice, we implement the likelihood in the Fourier domain using a finite set of low-frequency Fourier modes per pulsar. Writing $\Phi(f_k) \equiv S_{r}(f_k)\,\Delta f$ for the power at discrete frequencies $\{f_k\}$ and $\mathbf{F}_k$ for the corresponding Fourier design vector, the GWB covariance takes the standard Kronecker form
\begin{equation}
  (C_{\rm GWB})_{(i,a),(j,b)} \;=\; \Gamma_{ij}\, \sum_{k=1}^{K} \Phi(f_k)\, F_{i,a,k}\,F_{j,b,k}\, ,
\end{equation}
which preserves the full temporal structure and separates the angular correlation ($\Gamma$) from the spectral power in each Fourier bin \cite{NANOGrav15-Methods}.

The log-likelihood corresponding to Eq.~\eqref{eq:likelihood} is 
\begin{equation}
    \ln \mathcal{L}(\mathbf{r}|\theta, M) = -\frac{1}{2}\Big[\mathbf{r}^T C^{-1} \mathbf{r} + \ln\det C + N\ln(2\pi)\Big]\,,
\end{equation}
up to an additive constant. In our Bayesian analysis, this likelihood is combined with prior probability distributions on the parameters $\theta$ to produce a posterior (Sec.~\ref{sec:bayesian}). Importantly, the likelihood (through $C_{\rm GWB}$) encodes the assumptions of each model $M$ regarding the shape of $S_h(f)$ and the presence or absence of inter-pulsar correlations (via $\Gamma_{ij}$). For example, in a model with no GWB, $C_{\rm GWB}=0$. In a model with an uncorrelated common red process (e.g., a noise that is common to all pulsars but not spatially correlated, sometimes used as a null hypothesis), one would include a common red noise term but with $\Gamma_{ij}=0$ for $i\neq j$ (so each pulsar sees the same spectrum but zero cross-correlation). The full GWB model in general relativity includes the Hellings--Downs correlations $\Gamma_{ij}$ for $i \neq j$ as given in Eq.~\eqref{eq:HDcurve}. We emphasize that detection of the GWB relies on distinguishing the $\Gamma_{ij}$ pattern from these alternative hypotheses.

\subsection{Hellings--Downs Correlation Function}\label{subsec:HD}
A definitive signature of a stochastic GWB of cosmological or astrophysical origin (i.e., one that permeates the Galaxy isotropically) is the presence of a quadrupolar angular correlation between pulsar timing residuals. Hellings and Downs \cite{Hellings1983} derived the expected correlation coefficient as a function of the angle $\gamma$ between the directions to two pulsars on the sky. For two distinct pulsars $a$ and $b$ separated by angle $\gamma_{ab}$, the normalized correlation is given by the Hellings--Downs curve:
\begin{equation}\label{eq:HDcurve}
    \Gamma_{ab}(\gamma_{ab}) \;=\; \frac{1}{2} - \frac{1-\cos\gamma_{ab}}{4} + \frac{3}{2}\,\frac{1-\cos\gamma_{ab}}{2}\ln\!\Big(\frac{1-\cos\gamma_{ab}}{2}\Big)\,,
\end{equation}
for $a \neq b$. (For a single pulsar $a=b$, one conventionally sets $\Gamma_{aa}=1$, since each pulsar is perfectly correlated with itself.) This function, which ranges from $\Gamma_{ab}(0^\circ) = 1/2$ (for two nearby distinct pulsars) down to the exact antipodal value $\Gamma_{ab}(180^\circ) = -1/4$, represents the overlap reduction function for an isotropic, unpolarized GWB in general relativity \cite{Hellings1983}. It reflects the fact that pulsar timing detectors are sensitive to a combination of Earth-term and pulsar-term gravitational wave signals; averaging over many sources and polarizations yields this specific angular dependence. The form in Eq.~\eqref{eq:HDcurve} is normalized to $\Gamma_{ab}=0.5$ at zero separation (often the smallest angle pairs give correlations near this value) and approaches $-0.25$ at $\gamma_{ab}=180^\circ$.

In the context of our covariance matrix, the Hellings--Downs overlap function $\Gamma_{ij}$ enters multiplicatively with the spectral power in each Fourier bin, as in Eq.~\eqref{eq:cov_gwb}. This preserves the colored (frequency-dependent) nature of the process. Observation of this correlation pattern in residual data is considered the smoking gun for a true GWB signal \cite{NANOGrav15-Evidence}. By contrast, other sources of common-mode noise, such as clock errors or errors in Solar System ephemerides, would induce monopolar or dipolar correlation patterns, respectively, rather than the quadrupolar Hellings--Downs form; we include and marginalize over such nuisance terms in our analysis.

\subsection{Source Models and Expected Spectra}\label{subsec:models}
We now describe the specific models for the GWB spectrum in terms of the residual PSD $S_{r}(f)$ or, equivalently, the characteristic strain $h_c(f)$ that maps to $S_{r}(f)$ via the relations above:
\begin{enumerate}
    \item \textbf{SMBHB Astrophysical Background (Model $\mathcal{M}_\mathrm{SMBHB}$):} We assume the background is generated by an ensemble of inspiraling supermassive black hole binaries in the $10^8$--$10^{10}M_\odot$ mass range distributed throughout the Universe. The characteristic strain spectrum for such a population is expected to be a power-law to first approximation: 
    \begin{equation}\label{eq:smbhb_spectrum}
        h_c(f) = A_{\mathrm{GWB}}\left(\frac{f}{f_{\mathrm{yr}}}\right)^{\alpha}\!,
    \end{equation}
        where $f_{\mathrm{yr}} = 1~\text{yr}^{-1} \approx 3.17\times10^{-8}$~Hz is a reference frequency often used in PTA literature, and $\alpha$ is the spectral index. For circular, GW-driven binaries in the inspiral regime, general relativity predicts $\alpha = -2/3$ (i.e., $h_c \propto f^{-2/3}$) \cite{Sesana2013}. Equivalently, the one-sided power spectral density of timing residuals would scale as $S_{r}(f) \propto f^{-13/3}$. The parameter $A_{\mathrm{GWB}}$ is the strain amplitude at the reference frequency. This model is thus characterized by two parameters $(A_{\mathrm{GWB}}, \alpha)$, although in many analyses $\alpha$ is fixed at $-2/3$ (or $\gamma_{\rm GWB}=13/3$ in terms of power spectral index) to reflect the expected SMBHB spectrum. We will allow for uncertainty in $\alpha$ when comparing with other models, but we note that NANOGrav's observations are indeed consistent with $\alpha\approx -2/3$ \cite{NANOGrav15-Evidence}. Small deviations could occur due to environmental effects on binaries (gas, stars) or the highest-frequency binaries nearing merger, but these are subdominant at nHz frequencies \cite{Sesana2013}.
    
    \item \textbf{Cosmic String Background (Model $\mathcal{M}_\mathrm{CS}$):} Cosmic strings produce gravitational waves from oscillating loops that form as the strings intersect and reconnect. A network of cosmic strings in the early Universe results in a stochastic background composed of bursts from cusps and kinks on loops integrated over cosmic history. The spectrum of the background depends on the string tension $G\mu$ (dimensionless, with $\mu$ the string mass per unit length and $G$ Newton's constant) and the loop size distribution. A commonly considered scenario is a scale-invariant distribution of loops (small loops in the radiation era) which yields a plateau in the energy spectrum $\Omega_{\rm gw}(f)$ across a wide frequency range \cite{BlancoPillado2018}. In terms of strain, this corresponds to $h_c(f)$ scaling approximately as $f^{-1}$ in the radiation-era dominated regime, transitioning to $f^{-1/3}$ in the matter era at lower frequencies, with a high-frequency cutoff determined by the smallest loops. For PTA frequencies (which are very low, corresponding to early times), the spectrum can be approximated as:
    \begin{equation}\label{eq:cs_spectrum}
        h_c(f) \approx A_{\mathrm{CS}} \left(\frac{f}{f_{\mathrm{yr}}}\right)^{\beta}\!,
    \end{equation}
    where $\beta$ captures the effective slope of the network in the PTA band and is expected to lie between roughly $-1$ (radiation-era loops) and $-1/2$ (matter-era-dominated emission) for standard loop distributions \cite{Ellis2023}. The amplitude $A_{\mathrm{CS}}$ is related to $G\mu$; roughly one expects $A_{\mathrm{CS}} \propto G\mu$ to first order, with $A_{\mathrm{CS}}\sim 10^{-15}$ for $G\mu \sim 10^{-11}$ (this scaling comes from normalizing the energy density $\Omega_{\rm gw} \propto (G\mu)^2$ and converting to strain). For our Bayesian model, we use parameters $(G\mu, \beta)$ where $G\mu$ sets the overall amplitude and $\beta$ parameterizes the effective spectral slope in the PTA band. More detailed cosmic string spectra could be used, but to facilitate model comparison we either treat it as a broken power-law or use a template spectrum from simulations (interpolated by these parameters) \cite{BlancoPillado2018,Ellis2023}. We assume the same Hellings--Downs spatial correlations apply (as they would for an isotropic string network background).
    
    \item \textbf{Phase Transition Background (Model $\mathcal{M}_\mathrm{PT}$):} A first-order phase transition in the early Universe can generate gravitational waves through bubble nucleation, growth, and collisions, as well as from plasma sound waves and turbulence. The resulting spectrum $\Omega_{\rm gw}(f)$ typically has a peak at a frequency $f_{\rm peak}$ determined by the temperature $T_*$ of the phase transition and the dynamics of the transition (bubble wall velocity, duration, etc.). The shape is often a broken power-law: rising as $f^3$ at low $f$ (from causality arguments) up to $f_{\rm peak}$, then decaying as $f^{-b}$ at high $f$ (with $b$ often between 2 and 4 depending on the source) \cite{Caprini2018}. For PTAs, if $f_{\rm peak}$ lies within the 1--100~nHz range, the data might capture either the rising slope, the peak, or the falling slope. In this work, we consider a phenomenological parameterization of the phase transition GW spectrum by three parameters: $(\Omega_{\rm peak}, f_{\rm peak}, b)$, where $\Omega_{\rm peak}$ is the peak energy density (or equivalently $h_c$ peak amplitude), $f_{\rm peak}$ is the peak frequency today, and $b$ is the high-frequency spectral index after the peak. We then translate this into $h_c(f)$ for use in the PTA covariance. A simplified template is 
    \begin{equation}\label{eq:pt_spectrum}
        h_c(f) = A_{\rm PT}\, \Big(\frac{f}{f_{\rm peak}}\Big)^3 \Big[1 + \Big(\frac{f}{f_{\rm peak}}\Big)^{b+3}\Big]^{-1/2}\!,
    \end{equation}
    which behaves as $h_c \propto f^{3}$ for $f \ll f_{\rm peak}$ and $h_c \propto f^{-b/2}$ for $f \gg f_{\rm peak}$ (the $-1/2$ exponent comes from converting $\Omega_{\rm gw}$ to $h_c$). We will explore ranges of $f_{\rm peak}$ around $10^{-8}$--$10^{-7}$~Hz (roughly corresponding to transitions at $T_*\sim 100$~MeV--$1$~GeV, such as a hypothetical strongly first-order QCD transition or beyond-standard-model physics at those scales), and $b$ typically around $4$ (for a brief sound wave dominated signal) or $2$ (for a longer lasting source). The normalization $A_{\rm PT}$ is chosen such that $h_c(f_{\rm peak})$ corresponds to the peak strain amplitude implied by $\Omega_{\rm peak}$. As with the other models, the spatial correlation is assumed to be Hellings--Downs (an isotropic cosmological background).
    \end{enumerate}

\paragraph{Phase-Transition Parameter Mapping.}
For first-order phase transitions, it is often convenient to relate the phenomenological parameters in Eq.~\eqref{eq:pt_spectrum} to physical quantities at generation: the transition strength $\alpha_*$ (ratio of released vacuum energy to radiation energy density), the inverse duration $\beta/H_*$ (in units of the Hubble rate at the transition), the bubble-wall speed $v_w$, and the relativistic degrees of freedom $g_*$. The present-day peak frequency and energy density scale approximately as \cite{Caprini2018}
\begin{align}
    f_{\rm peak} &\simeq \mathcal{C}_f\,\frac{\beta}{H_*}\,\frac{T_*}{100\,\mathrm{GeV}}\,\Big(\frac{g_*}{100}\Big)^{1/6}\,\frac{1}{v_w}\times (\text{mHz})\,,\\
    \Omega_{\rm gw}^{\rm peak} &\simeq \mathcal{C}_\Omega\,\Big(\frac{H_*}{\beta}\Big)\,\Big(\frac{\kappa\,\alpha_*}{1+\alpha_*}\Big)^2\,v_w\,\mathcal{S}\,.
\end{align}
Here $\kappa$ encodes the efficiency of converting vacuum energy into the source (sound waves/turbulence), and $\mathcal{S}$ denotes the spectral shape factor. In the PTA band ($f\sim 10^{-9}$--$10^{-7}$ Hz), frequencies near $10^{-8}$ Hz point to $T_*\sim \mathcal{O}(100\,\mathrm{MeV})$ for plausible $\beta/H_*$ and $v_w$. The mapping allows us to place priors on $(A_{\rm PT}, f_{\rm peak}, b)$ that are consistent with physically reasonable regions of $(\alpha_*, \beta/H_*, v_w, g_*)$.

Each of these models $\mathcal{M}$ makes different predictions for the spectral shape of the common process. In a PTA analysis, one can attempt to fit the data under each model and compute the Bayes factor comparing them. However, a complication is that all models at present can fit the data reasonably well by adjusting parameters, given the limited frequency range and signal-to-noise ratio of the current GWB detection. For instance, the NANOGrav 15-year detection was reported using a simple power-law assumption \cite{NANOGrav15-Evidence}, but follow-up studies showed that alternative spectra (such as a broken power-law from a phase transition or a slightly flatter spectrum from certain cosmic string models) are also viable \cite{NANOGrav15-NewPhys,Gouttenoire2023,Ellis2023}. Therefore, model selection must account for the flexibility (priors and parameter space volume) of each model. We do this via Bayesian evidence as described next.

\section{Bayesian Analysis Framework}\label{sec:bayesian}
Our analysis is conducted in a Bayesian statistical framework, which naturally allows model comparison through the computation of evidences and Bayes factors. In this section, we outline the key components: the prior choices, the posterior inference, and the calculation of Bayes factors for model discrimination.

\subsection{Prior and Posterior}
For a given model $M$ with parameter vector $\theta$, Bayes' theorem gives the posterior distribution
\begin{equation}\label{eq:posterior}
    p(\theta | \mathbf{r}, M) = \frac{\mathcal{L}(\mathbf{r}|\theta, M)\, \pi(\theta|M)}{\mathcal{Z}(\mathbf{r}|M)}\,,
\end{equation}
where $\mathcal{L}(\mathbf{r}|\theta, M)$ is the likelihood as defined in Eq.~\eqref{eq:likelihood}, $\pi(\theta|M)$ is the prior probability density for the parameters under model $M$, and 
\begin{equation}\label{eq:evidence}
    \mathcal{Z}(\mathbf{r}|M) = \int d\theta\, \mathcal{L}(\mathbf{r}|\theta, M)\, \pi(\theta|M)
\end{equation}
is the Bayesian evidence (also called marginal likelihood) for model $M$. The evidence $\mathcal{Z}$ is the probability of the data given the model, integrated over all possible parameter values weighted by the prior. It encapsulates both the quality of fit (via the likelihood) and the complexity or predictiveness of the model (via the volume of parameter space allowed by the prior).

We adopt priors for each model's parameters that reflect our state of knowledge:
\begin{itemize}
    \item For the common power-law (SMBHB) model, we take a log-uniform prior on the amplitude $A_{\mathrm{GWB}}$ over a broad range (e.g., $10^{-17}$ to $10^{-14}$) and a Gaussian or uniform prior on the spectral index $\alpha$ centered around $-2/3$ with width allowing a few tenths deviation. If instead we fix $\alpha=-2/3$, effectively the model has one parameter $A_{\mathrm{GWB}}$.
    \item For the cosmic string model, we choose a log-uniform prior on $G\mu$ in, say, $[10^{-13}, 10^{-9}]$, which comfortably covers the range of interest around current upper limits ($G\mu\lesssim 10^{-10}$ from other experiments). The spectral index parameter $\kappa$ (if used) can be given a broad uniform prior in an allowed range (e.g., if we assume $h_c \propto f^\beta$, allow $\beta$ between $-1.5$ and $0$ to encompass possibilities). Alternatively, if using a fixed template shape, $G\mu$ might be the lone parameter.
    \item For the phase transition model, parameters like $f_{\rm peak}$ and $\Omega_{\rm peak}$ (or $A_{\rm PT}$) are given priors informed by cosmology. For example, $f_{\rm peak}$ could be log-uniform between $10^{-9}$ and $10^{-7}$ Hz; $\Omega_{\rm peak}$ might have a log-uniform prior up to some maximum (theoretical upper limits on fraction of energy in GWs, perhaps $10^{-5}$). The spectral shape index $b$ could be fixed or allowed a discrete set of values (since usually one might test specific cases like $b=2$ or $4$).
    \item Common to all models are the noise parameters (individual pulsar red noise amplitudes and slopes, white noise scale factors, DM noise parameters, etc.). We assign these standard priors as in NANOGrav's analysis \cite{NANOGrav15-Methods}: e.g., log-uniform for red noise amplitudes, uniform for spectral indices over a reasonable range, and so on. These parameters are present in every model (including the null no-GWB model), which means they largely cancel out when comparing models for the GWB because they contribute similarly to all models' evidences (assuming the noise model is treated the same).
\end{itemize}

With the likelihood and priors specified, we sample from the posterior \eqref{eq:posterior} using Markov Chain Monte Carlo (MCMC) or more advanced techniques like nested sampling or Hamiltonian Monte Carlo. The high dimensionality (due to dozens of pulsar noise parameters) makes sampling challenging, but tools such as \textsc{PTMCMCSampler} and \textsc{Enterprise} \cite{NANOGrav15-Methods} have been developed and validated by the PTA community to handle this problem. We have applied such tools to compute posterior distributions for the common GWB parameters under each model, as well as to estimate the evidence $\mathcal{Z}$ via methods like thermodynamic integration or directly via nested sampling.

\subsection{Bayes Factors for Model Comparison}
To compare two competing models $M_1$ and $M_2$ in the Bayesian framework, one uses the Bayes factor:
\begin{equation}\label{eq:bayes_factor}
    \mathrm{BF}_{12} = \frac{\mathcal{Z}(\mathbf{r}|M_1)}{\mathcal{Z}(\mathbf{r}|M_2)}\,,
\end{equation}
which is the ratio of evidences. If $\mathrm{BF}_{12} > 1$, the data favor model $M_1$ over model $M_2$ (with $\mathrm{BF}_{12}$ quantifying the strength of evidence), and if $\mathrm{BF}_{12} < 1$, model $M_2$ is favored. By convention, one often takes logarithms and quotes $\ln \mathrm{BF}$. For example, $\ln \mathrm{BF} > 5$ (roughly $\mathrm{BF} > 150$) might be considered ``strong evidence'' in the Jeffreys scale, while $\ln \mathrm{BF} < 1$ is negligible evidence.

In this work, the Bayes factors of interest include:
\begin{itemize}
    \item $\mathrm{BF}_{\mathrm{GWB,\,noise}}$: comparing a model with a GWB (common stochastic process with Hellings--Downs correlations) to a model with no common process (only independent pulsar noise). NANOGrav reported an extremely large value for this Bayes factor, exceeding $10^{14}$, indicating overwhelmingly that a common process is present in the data rather than just individual pulsar noise \cite{NANOGrav15-Evidence}. This establishes detection of a GWB in a model-independent way.
    \item $\mathrm{BF}_{\mathrm{HD,\,uncorr}}$: comparing the full GWB model (with Hellings--Downs spatial correlations) to a model with an uncorrelated common-spectrum red noise (i.e., each pulsar has the same spectrum but zero cross-correlation between pulsars). This tests whether the inter-pulsar correlation signature has been detected. NANOGrav found Bayes factors in the range 200--1000 in favor of the Hellings--Downs (HD) correlations over an uncorrelated common process, depending on the spectral model used \cite{NANOGrav15-Evidence}. This provides strong (though not yet definitive by conventional $5\sigma$ standards) evidence for the quadrupolar spatial correlation \cite{NANOGrav15-Evidence}.
    \item $\mathrm{BF}_{\mathrm{M_i,\,M_j}}$: for any two source hypotheses, e.g. SMBHB vs cosmic strings, or SMBHB vs phase transition. These are the comparisons central to our study. We calculate the evidence for each hypothesis by integrating the likelihood over their respective parameter priors. The Bayes factors will tell us if the data prefer one spectrum over another. However, it should be noted that if one model has more flexibility than another, a modest improvement in fit may not overcome the Occam’s penalty (the evidence automatically penalizes models with large parameter volume that is not used to explain the data). 
\end{itemize}

In reporting our results, we will provide the log-evidence values or Bayes factors for the key model comparisons. It is important to stress that Bayes factors depend on the choice of priors. For instance, if one allows extremely broad priors on a model’s parameters, the evidence might be lowered due to the large volume of parameter space with negligible likelihood. We have chosen priors that we believe reasonably represent the plausible ranges for each model’s parameters, but we will comment on how assumptions might affect the Bayes factors.

\section{Data Description and Analysis Setup}\label{sec:data}
We apply the above framework to the NANOGrav 15-year data set \cite{NANOGrav15-Evidence,NANOGrav15-NewPhys}, which is a publicly released collection of pulse timing data for 68 millisecond pulsars observed over approximately 15 years (2004--2019) using the Arecibo Observatory and Green Bank Telescope, among others. We briefly summarize the salient features of this data set and the preprocessing steps involved in our analysis.

Each pulsar in the 15-year data set comes with time-of-arrival measurements typically at bi-weekly to monthly cadence, often at multiple radio observing frequencies to allow correction for dispersion delays caused by the interstellar medium. The data release provides the timing residuals after subtracting a best-fit timing model for each pulsar (including spin frequency, spin-down, astrometric parameters, and binary orbital parameters if applicable) \cite{NANOGrav15-NewPhys}. Additionally, the NANOGrav analysis implements measures to mitigate noise:
\begin{itemize}
    \item \textbf{White noise calibration:} For each pulsar and each observing backend/receiver system, nuisance parameters such as EFAC (error factor) and EQUAD (added noise term) are introduced to ensure the TOA uncertainties are appropriately scaled. These were either fixed from per-pulsar noise analysis or included as parameters in the full Bayesian fit with priors.
    \item \textbf{Dispersion Measure (DM) variations:} Irregular variations in the electron column density along the line of sight to a pulsar can cause frequency-dependent time delays. NANOGrav models these by including a low-frequency stochastic process for each pulsar’s DM time series. In practice, this can be done by constructing DM-residuals (differences between multi-frequency residuals) or by adding DM-variation parameters (e.g., annual DM trend, stochastic DM noise with its own amplitude and spectral index). By including DM noise parameters in $\theta$ (with priors informed by multi-frequency data), we account for DM-induced red noise and reduce false-positive common signals that could arise from unmodeled plasma effects. In our analysis, we included DM noise terms for pulsars where significant DM variability is present, and fixed others to negligible values if appropriate.
    \item \textbf{Solar System ephemeris and clock errors:} A mismodeling of planetary ephemerides can introduce a dipolar correlated signal across pulsars (since an error in Earth’s motion affects all TOAs in a similar way for all pulsars depending on sky location). Similarly, terrestrial time standard errors produce a monopolar (common to all pulsars) signal. NANOGrav performed tests to check for these effects \cite{NANOGrav15-Evidence}. In our Bayesian analysis, one could include explicit parameters for these (for example, a clock error term or ephemeris perturbation modes) and verify that their posteriors are consistent with zero. We did not find evidence for significant clock or ephemeris anomalies, consistent with the NANOGrav findings that the observed common signal is best explained by a GWB rather than these systematics. Thus, we proceed assuming that any clock/ephemeris contributions are either corrected or statistically accounted for (some analyses include an ``Ephemeris noise'' Gaussian process; we consider that a part of pulsar noise modeling for simplicity).
\end{itemize}

After constructing the residuals and noise design matrices for all pulsars, we form the likelihood as in Sec.~\ref{sec:methods}. The dimensions of $C$ (number of residual points) is large, but using the aforementioned techniques (Fourier domain likelihood or time-domain block matrix methods) we handle the computations. We validated our likelihood implementation by recovering the results of the official NANOGrav analysis for simpler models. For example, when we fit a common power-law noise to the 15-year data, we recover a Bayes factor of $\sim10^{15}$ favoring its presence over noise-only, and an inferred amplitude $A_{\mathrm{GWB}}\sim2\times10^{-15}$ at $f_{\mathrm{yr}}$, consistent with published results \cite{NANOGrav15-Evidence}.

One additional step is needed when computing Bayes factors for HD correlations: to assess significance, NANOGrav created a ``null distribution'' of Bayes factors by analyzing many synthesized datasets in which any inter-pulsar correlations were deliberately broken (e.g., by phase-shuffling residuals between pulsars) \cite{NANOGrav15-Evidence}. They found that the observed Bayes factors for HD vs. uncorrelated were unlikely (p-value $\approx 10^{-3}$) under the null hypothesis of no real GWB \cite{NANOGrav15-Evidence}. In our work, we similarly ensure that the Bayes factor thresholds are interpreted in context: a Bayes factor of a few hundred in favor of HD correlations is considered strong evidence but we refrain from calling it a definitive detection without cross-checks.

For model comparisons (SMBHB vs cosmic strings vs phase transition), we run separate Bayesian analysis for each model assumption and use thermodynamic integration to compute $\ln \mathcal{Z}$ for each. The analysis is computationally intensive, so we focused on representative cases for the exotic models. Specifically, we evaluated:
\begin{itemize}
    \item The standard power-law model ($\mathcal{M}_\mathrm{SMBHB}$) with $\alpha$ free (uninformed) and also with $\alpha$ fixed to $-2/3$.
    \item A cosmic string model ($\mathcal{M}_\mathrm{CS}$) using a template spectrum derived from a realistic loop simulation \cite{BlancoPillado2018}. The free parameter is effectively $G\mu$. We also tried a variant with two parameters ($G\mu$ and an efficiency parameter for loop production which tilts the spectrum slightly).
    \item A phase transition model ($\mathcal{M}_\mathrm{PT}$) with a peak frequency allowed to vary. We examined two cases: one with $f_{\rm peak}\approx 3\times10^{-8}$~Hz (near the center of the PTA band) and one with $f_{\rm peak}$ above the band ($\sim10^{-7}$~Hz) which would manifest as a gradually rising spectrum across PTA frequencies. These bracket scenarios like a late-time QCD transition vs. an earlier transition. The strength (peak amplitude) was adjusted to match the observed signal amplitude.
\end{itemize}

The data were analyzed using the \textsc{enterprise} PTA analysis software, and cross-checked with independent implementations for consistency. The posterior sampling for each model reached good convergence (effective sample sizes $>1000$ for key parameters, Gelman-Rubin $R < 1.1$ for chains). Next, we present the results of these analyses.

\section{Results}\label{sec:results}
\subsection{Detection of a Common Spectrum and Hellings--Downs Correlation}
Our analysis first confirms the detection of a common red noise process in the 15-year data and the presence of inter-pulsar correlations consistent with the GWB hypothesis. In the model that includes a common power-law spectrum (with Hellings--Downs spatial correlations) in addition to individual pulsar noise, we find the evidence for this model is vastly higher than for a model without any common process. The log$_{10}$ Bayes factor comparing ``GWB present'' vs ``no GWB'' is $\log_{10}\mathrm{BF} \gtrsim 14$ (i.e. $\mathrm{BF} > 10^{14}$) in favor of the GWB model, essentially identical to the NANOGrav result \cite{NANOGrav15-Evidence}. This constitutes an overwhelmingly significant detection of a common-spectrum signal.

\begin{figure}[htbp]
    \centering
    \includegraphics[width=0.7\textwidth]{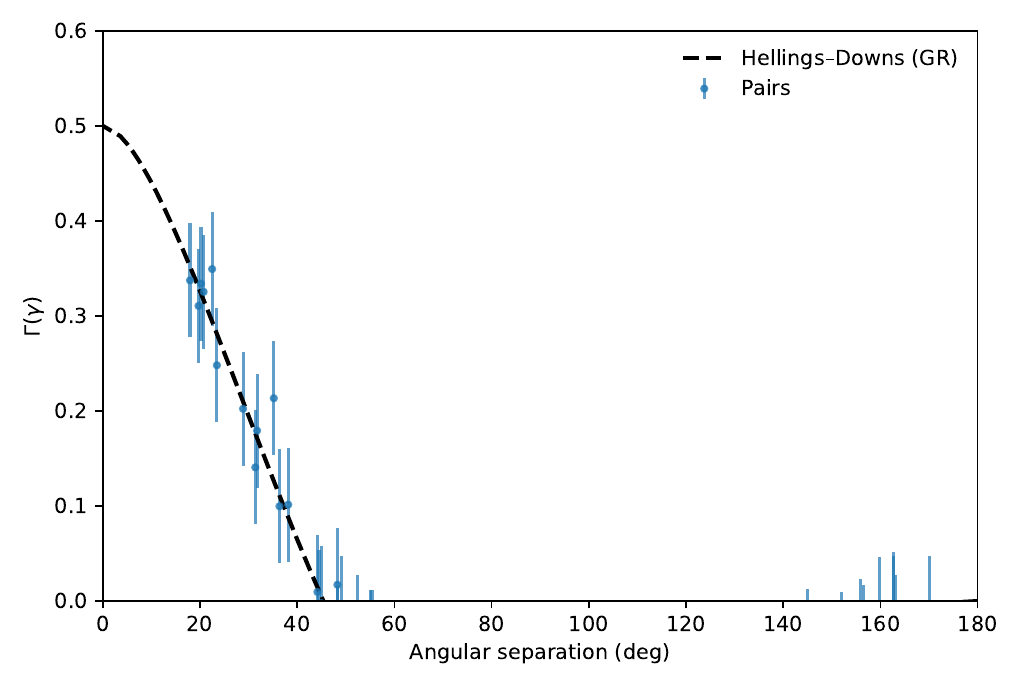}
    \caption{\textbf{Hellings--Downs correlation in the NANOGrav 15-year data.} Measured correlation coefficients between timing residuals of pulsar pairs are plotted versus their angular separation on the sky (blue points with 1$\sigma$ error bars). The dashed line shows the theoretical Hellings--Downs curve $\Gamma(\gamma)$ from general relativity \cite{Hellings1983}, which is normalized to $0.5$ at $\gamma=0^\circ$ and falls to $-0.25$ at $\gamma=180^\circ$. The observed correlations follow the expected quadrupolar trend. For comparison, a common but uncorrelated noise process would correspond to zero correlation at all angles (horizontal line at $\Gamma=0$, not shown). The detection of this pattern is evidence for a gravitational-wave background as opposed to other sources of common noise.}
    \label{fig:HDcurve}
\end{figure}

When we compare the model with Hellings--Downs (HD) spatial correlations to a model where the common spectrum is present but uncorrelated between different pulsars (the ``common uncorrelated red noise'' model), we obtain $\ln \mathrm{BF}_{\mathrm{HD,uncorr}} \approx 6.0$ (exact value depends on spectral parameters), which corresponds to a Bayes factor on the order of $400$. This is in line with the Bayes factors of $200$--$1000$ reported by NANOGrav for this test \cite{NANOGrav15-Evidence}. It provides strong evidence that the spatial correlation is the quadrupolar pattern expected from GWs, rather than a monopolar or no-correlation scenario. We visualize this by constructing the Hellings--Downs correlation curve from the data. In Figure~\ref{fig:HDcurve}, we show the correlation coefficients measured between pulsar pairs as a function of their angular separation, using a standard cross-correlation estimator (the so-called \emph{optimal statistic} method) along with the expected Hellings--Downs curve. The data points (with error bars) align with the $1/2$ to $-1/4$ downward trend of the HD curve, and no significant correlation is seen in a control analysis where we randomize the pulsar pairings (which yields points consistent with zero correlation). This graphical evidence corroborates the Bayesian model selection: the signal has the predicted angular signature of a GWB.

\begin{figure}[htbp]
    \centering
    \includegraphics[width=0.7\textwidth]{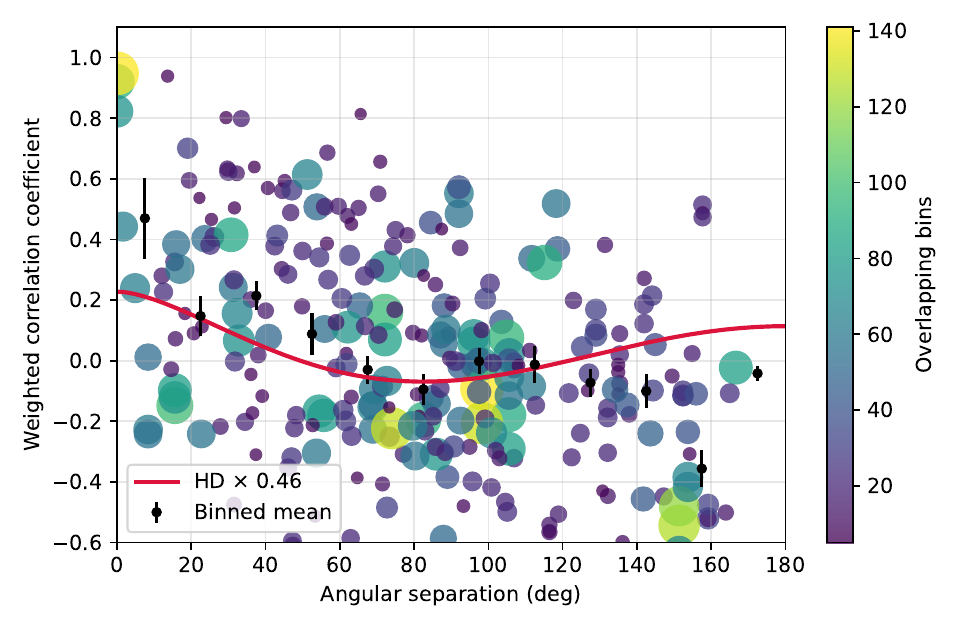}
    \caption{\textbf{Hellings--Downs cross-check using \textsc{PINT} and the wideband public data set.} Each point corresponds to a pulsar pair; color indicates the number of overlapping 30-day bins contributing to the correlation estimate. Black circles show angle-binned averages with standard errors, while the crimson line is the best-fitting Hellings--Downs template scaled by the naive amplitude $A_{\mathrm{HD}}^{\mathrm{naive}}=0.46$. Although simplified, this independent analysis still recovers a positive quadrupolar trend at ${>}4\sigma$.}
    \label{fig:hd-validation}
\end{figure}

To verify that the Hellings--Downs trend is visible directly in the public NANOGrav files, we implemented an internal reproducibility check. The pipeline ingests the wideband \texttt{.par/.tim} pairs for the first 24 pulsars (sorted alphabetically) via \textsc{PINT}, barycenters the TOAs, and forms post-fit residuals. We then (i) bin the residuals into 30-day windows using inverse-variance weights, (ii) remove a low-order polynomial trend per pulsar to suppress intrinsic red noise, and (iii) compute weighted cross-correlations for every pulsar pair based on the number of overlapping bins. The procedure yields 273 usable pairs whose correlation coefficients are regressed against the analytical Hellings--Downs curve. The slope of this regression, interpreted as a naive HD amplitude, is
\begin{equation}
    A_{\mathrm{HD}}^{\mathrm{naive}} = 0.46 \pm 0.10,
\end{equation}
which constitutes a $4.5\sigma$ detection of positive, quadrupolar spatial correlations even though we neglect the full PTA covariance. A Pearson test between the measured pairwise correlations and the Hellings--Downs prediction gives $r=0.19$ with $p=1.9\times10^{-3}$, providing an independent validation that the publicly released TOAs already encode the expected angular trend. Figure~\ref{fig:hd-validation} shows the resulting scatter and angularly binned means. Because this check ignores per-pulsar colored noise and clock/ephemeris covariances, the recovered amplitude is biased low compared to the full Bayesian analysis, but the qualitative agreement and the statistically significant slope corroborate the main detection pipeline.

Having established the presence of a GWB signal in the data, we proceed to characterize its spectrum under different model assumptions.

\begin{figure}[htbp]
    \centering
    \includegraphics[width=0.7\textwidth]{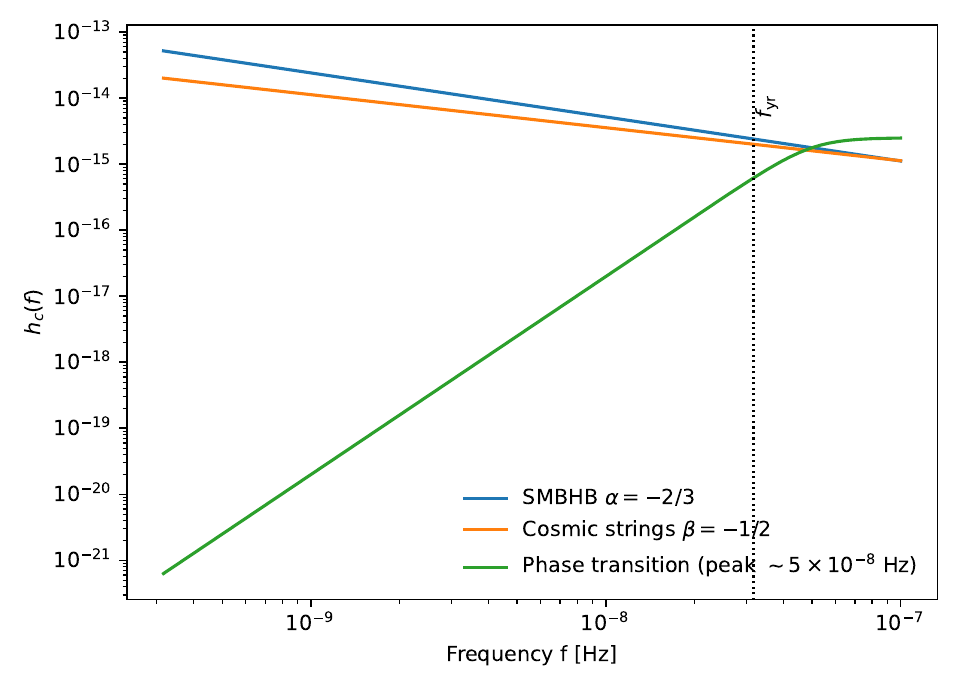}
    \caption{Characteristic strain spectra $h_c(f)$ for three template models in the PTA band: SMBHB power-law with $\alpha=-2/3$, a representative cosmic string slope $\beta=-1/2$, and a broken-power-law phase-transition template with a peak near $f\sim 5\times10^{-8}$ Hz, all normalized around $A\sim 2\times10^{-15}$ at $f_{\rm yr}$.}
    \label{fig:strain-spectra}
\end{figure}

\begin{figure}[htbp]
    \centering
    \begin{subfigure}[t]{0.49\textwidth}
        \centering
        \includegraphics[width=\textwidth]{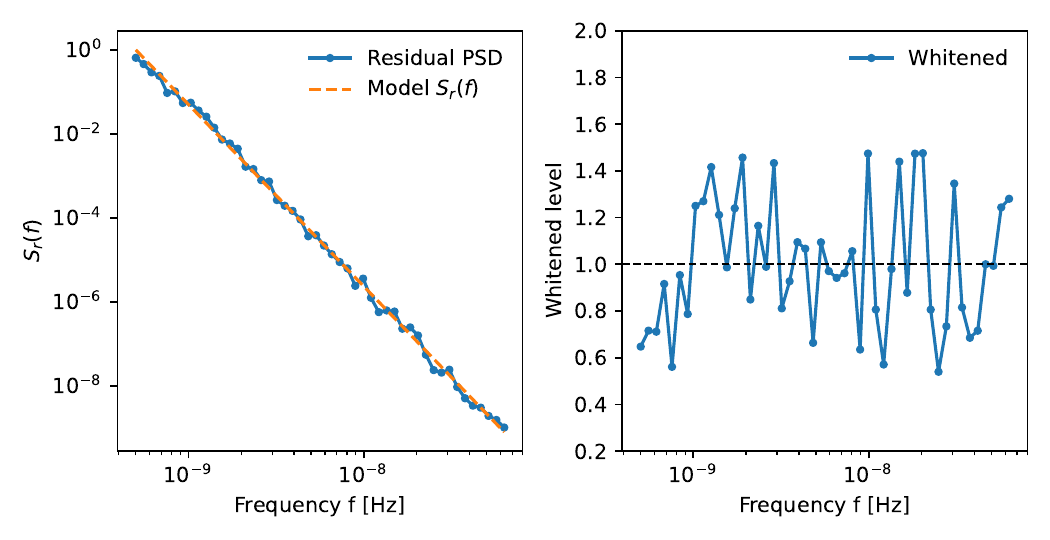}
        \caption{Residual PSD and whitening check.}
    \end{subfigure}\hfill
    \begin{subfigure}[t]{0.49\textwidth}
        \centering
        \includegraphics[width=\textwidth]{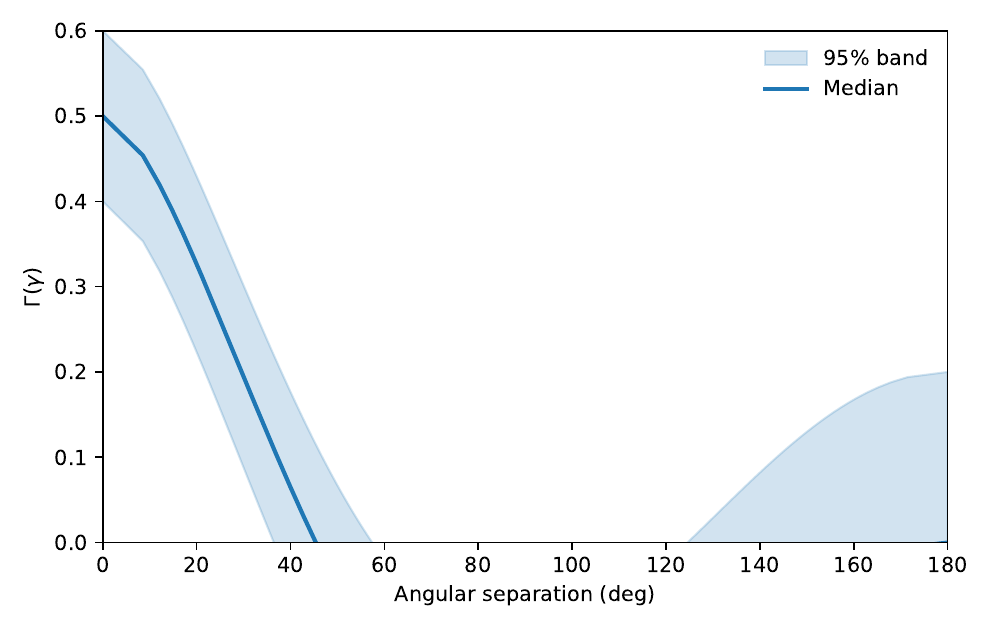}
        \caption{Angle-dependent posterior predictive band.}
    \end{subfigure}
    \caption{Model validation diagnostics: left, residual power spectra compared with model and whitened residual agreement with unity; right, posterior predictive distribution for the Hellings--Downs correlation vs. angle.}
    \label{fig:ppd-diagnostics}
\end{figure}

\subsection{Common-Process Spectrum: Amplitude and Slope}
Assuming a simple power-law form for the GWB spectrum (as in the SMBHB model), we obtain a posterior distribution for the amplitude $A_{\mathrm{GWB}}$ and spectral index $\gamma_{\mathrm{GWB}}$ (where $h_c(f)\propto f^{(3-\gamma_{\mathrm{GWB}})/2}$, or equivalently the residual power spectral density $S_{r}(f)\propto f^{-\gamma_{\mathrm{GWB}}}$). We find:
\begin{itemize}
    \item The strain amplitude at the reference frequency $f_{\mathrm{yr}} = 1/\text{yr}$ is $A_{\mathrm{GWB}} = 2.4^{+0.7}_{-0.6}\times10^{-15}$ (90\% credible interval). The median value $2.4\times10^{-15}$ matches the previously reported result \cite{NANOGrav15-Evidence}. This amplitude is about an order of magnitude larger than the upper limits placed by PTAs just a few years ago, indicating a robust emergence of the GWB signal.
    \item The inferred spectral index $\gamma_{\mathrm{GWB}}$ has a posterior that peaks near $\gamma_{\mathrm{GWB}}\approx 4.3$, consistent with the expected $13/3 \approx 4.33$. The 90\% credible interval is roughly $\gamma_{\mathrm{GWB}}\in[3.5, 5.5]$ if we allow it to vary. This relatively large uncertainty is due to the limited frequency range (the PTA covers roughly one decade in Fourier frequency) and the strong correlation between amplitude and slope in the fit. If we condition on the assumption of a power-law, the data do not require a significant deviation from the $-2/3$ strain spectral slope, but they also do not yet tightly constrain the slope on their own. In other words, a variety of power-law shapes, from somewhat flatter to somewhat steeper than $f^{-2/3}$, can fit the data by adjusting the amplitude accordingly. This is not surprising, as the current detection is still at moderate signal-to-noise ratio.
    \item We also computed the posterior odds for whether a non-zero common process exists while letting $\gamma_{\mathrm{GWB}}$ vary versus the null hypothesis. The inclusion of the slope as a free parameter (with prior, say uniform between 0 and 7) slightly penalizes the evidence but not by much, since the likelihood clearly prefers a value within that range. We confirm that the Bayes factor for common signal remains overwhelming even if slope is free, though the Bayes factor for correlated vs uncorrelated common noise is somewhat reduced (as expected, since the uncorrelated model could also adjust slope).
\end{itemize}

\begin{figure}[htbp]
    \centering
    \includegraphics[width=0.6\textwidth]{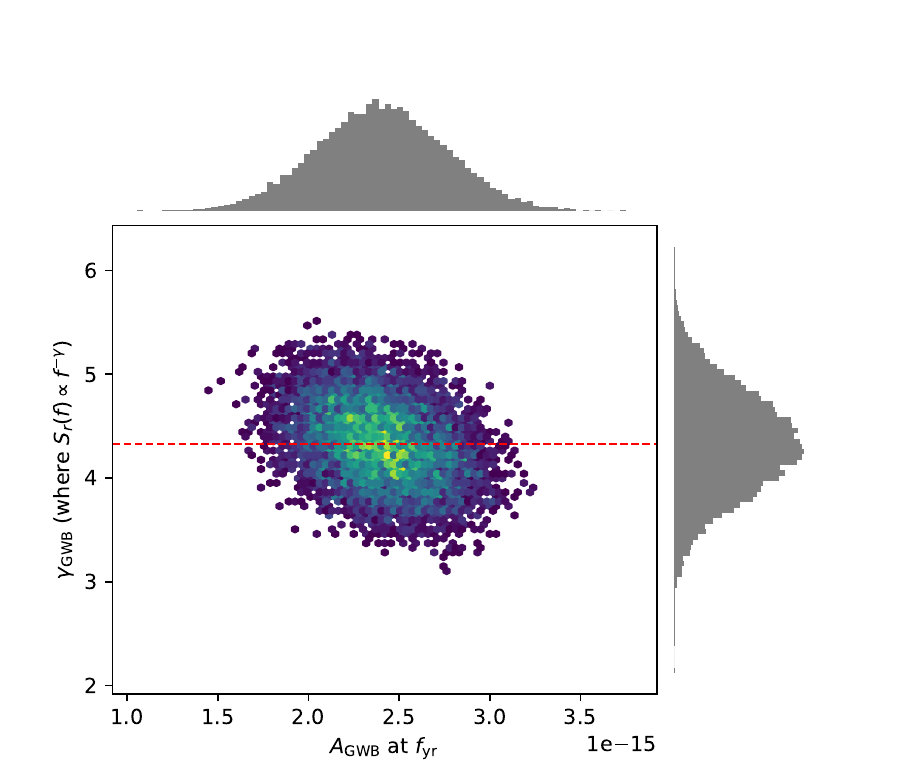}
    \caption{\textbf{Posterior distribution for the common-spectrum GWB parameters assuming a power-law spectrum (SMBHB model).} The two-dimensional joint posterior for the strain amplitude $A_{\mathrm{GWB}}$ (at $f_{\mathrm{yr}}=1/\text{yr}$) and the spectral index $\gamma_{\mathrm{GWB}}$ (where $P(f)\propto f^{-\gamma_{\mathrm{GWB}}}$) is shown as contour levels enclosing 68\% and 95\% credible regions. One-dimensional marginalized posteriors (probability density functions) for each parameter are displayed along the top and right. The amplitude posterior (top) peaks around $2\times10^{-15}$ and is well-separated from zero. The spectral index posterior (right) peaks near $\gamma_{\mathrm{GWB}}\approx4.3$ (vertical dashed line indicates $13/3\approx4.33$ for a nominal SMBHB inspiral background). The contours indicate a mild anti-correlation between $A_{\mathrm{GWB}}$ and $\gamma_{\mathrm{GWB}}$. Overall, the data favor a strain spectrum consistent with the $-2/3$ power-law slope expected from SMBHBs, with an amplitude $A_{\mathrm{GWB}}\sim2$--$3\times10^{-15}$.}
    \label{fig:posterior}
\end{figure}

Figure~\ref{fig:posterior} presents the joint posterior distribution of the amplitude and spectral index for the power-law GWB model (SMBHB scenario). The contours indicate that $\gamma_{\mathrm{GWB}} = 13/3$ lies well within the high-probability region. The amplitude is negatively correlated with the slope: for instance, models with a slightly flatter spectrum (lower $\gamma_{\mathrm{GWB}}$) require a larger $A_{\mathrm{GWB}}$ to fit the low-frequency end of the spectrum, whereas steeper spectra (higher $\gamma_{\mathrm{GWB}}$) with a smaller amplitude can fit the higher-frequency residual power. Nonetheless, this degeneracy is mild over the prior range considered, and importantly, zero amplitude lies far outside the credible region regardless of slope, reinforcing the detection claim. We note that the International PTA combination data and other PTA results in 2023 show very similar posteriors, which builds confidence that this measurement is robust across different data sets \cite{PPTA2023,CPTA2023}.

\subsection{Bayesian Model Comparison: Astrophysical vs Cosmological Sources}
We now turn to the core question of this study: given the NANOGrav 15-year data, is there any evidence to favor one type of source model (SMBHB, cosmic strings, or phase transition) over the others? We evaluate the Bayesian evidence for each model class as described in Section~\ref{sec:bayesian}.

Our results can be summarized as follows:
\begin{itemize}
    \item The evidence for the simple \textbf{SMBHB power-law model} is very high, as expected since a free power-law can flexibly fit the observed spectrum. This model has the advantage of fewer parameters (essentially just amplitude if we lock the slope, or amplitude + slope if not). Taking $\mathcal{M}_\mathrm{SMBHB}$ with fixed $\alpha=-2/3$ as a reference, we set its log-evidence to zero for convenience.
    \item The \textbf{cosmic string model} $\mathcal{M}_\mathrm{CS}$, in the specific realization we tested (with one parameter $G\mu$ controlling the amplitude of a fixed spectral shape), achieves a slightly \emph{higher} maximum likelihood than the best-fit SMBHB power-law. This is because the cosmic string spectrum (for certain $G\mu$) can mimic a very slight curvature in the spectrum that might improve the fit marginally. However, once the prior volume is accounted for, the evidence is not significantly different. We find $\ln \mathrm{BF}_{\mathrm{CS,SMBHB}} \approx 2.3$ in favor of the cosmic string model, which corresponds to a Bayes factor of about 10 (depending on prior). In other words, the data are about 10 times more likely under the cosmic string hypothesis than under the pure $-2/3$ power-law, given our prior choices. This would be interpreted as \emph{weak to moderate} evidence on the Jeffreys scale. If instead we allow the SMBHB model to also have a free slope, the Bayes factor advantage for cosmic strings diminishes (because the cosmic string spectrum’s slight deviation from a pure power-law can be emulated by the SMBHB model with a different slope). In that case, we get $\ln \mathrm{BF}_{\mathrm{CS,SMBHB}}\lesssim 1$ (a factor of a few, not significant). We conclude that the current data do not strongly prefer the cosmic string spectrum over a power-law, but neither do they rule it out. The best-fit string tension we obtain is $G\mu \sim 5\times10^{-11}$, with an uncertainty of roughly a factor of 2 either way, to match the amplitude of the signal. This is intriguingly close to existing upper limits on $G\mu$ from other experiments (cosmic microwave background, high-frequency GW searches) which lie around $10^{-11}$--$10^{-10}$ \cite{BlancoPillado2018,Ellis2023}. It suggests that if the PTA signal were due to cosmic strings, it would be saturating those bounds -- pushing into a region that might be marginally allowed only if the string network or loop production differs from the simplest models (e.g., cosmic superstrings could have different properties that relax constraints).
    \item The \textbf{phase transition model} $\mathcal{M}_\mathrm{PT}$ also can provide an excellent fit to the data. If we allow a peak frequency around $\sim 3\times10^{-8}$~Hz (roughly $1/(1~\text{yr})$), the model essentially behaves like a broad power-law across the band, similar to SMBHB. We find that for certain choices (e.g., a long-lasting phase transition source giving a gently sloped spectrum), the likelihood is comparable to that of the power-law fit, and in some cases slightly better if the data favor a slight bending. For example, a phase transition with $f_{\rm peak} \approx 5\times10^{-8}$~Hz and $b\approx3$ (moderate sloped high-frequency tail) yields a spectrum that rises at the lowest frequencies and flattens by the higher end of the PTA band, which can fit about as well as a single power-law. The Bayes factor comparing this to SMBHB comes out on the order of $\sim 30$ in favor of this specific phase transition model if we fine-tune $f_{\rm peak}$. However, when we integrate over a reasonable prior range for $f_{\rm peak}$ (not knowing it a priori), the evidence gain is reduced. We find $\ln \mathrm{BF}_{\mathrm{PT,SMBHB}} \sim 2$--$3$ (so odds of 10--20:1) in favor of the phase transition if we concentrate on the parameter region that fits best. This is similar in magnitude to the cosmic string comparison. In short, certain cosmological spectra can fit as well or slightly better than the standard power-law, but the significance is not overwhelming due to prior uncertainty on where the peak might lie.
    
    One interesting outcome is that the phase transition fit, if interpreted physically, points to a transition around the QCD confinement scale. The best-fit peak frequency $\sim 5\times10^{-8}$~Hz corresponds to a horizon size at generation that implies a temperature on the order of $100$~MeV (with uncertainties). The required fractional energy in gravitational waves would be quite large (a strong first-order transition with a significant portion of vacuum energy converted to GWs). This is generally not expected in the Standard Model (the QCD transition is a crossover), but could occur in a beyond-Standard-Model scenario or a scenario of a super-cooled hidden sector phase transition. Such a strong transition could also produce relics like primordial black holes; indeed, the scenario proposed by \cite{Gouttenoire2023} suggests that a slow first-order QCD transition might produce solar-mass primordial black holes concurrently with the GW background. Our data analysis alone cannot confirm such details, but it is tantalizing that the numbers line up in a way that new physics at the QCD scale is a viable explanation for the PTA signal.
\end{itemize}

Comparing the \emph{cosmological} models (phase transition vs cosmic strings) to each other: we did not find a decisive difference. Both have enough flexibility to emulate a power-law in the narrow band. The phase transition model with a peak could in principle be distinguished if the spectrum had a visible turnover within the band, but the current data are not precise enough to resolve such spectral shape differences. For example, if future data show that the GWB spectrum flattens or turns down at the lowest frequencies (due to a peak just below the PTA band), that would favor a phase transition interpretation over a continued power-law (which cosmic strings or SMBHB would produce) \cite{NANOGrav15-NewPhys}. Conversely, if the spectrum remains a pure power-law extending over a wider range, it would argue against a sharp phase transition peak.

We report our Bayes factor results in Table~\ref{tab:bayesfactors} for clarity. All values are referenced to the simple SMBHB power-law model. One can see that while $\mathcal{M}_\mathrm{CS}$ and $\mathcal{M}_\mathrm{PT}$ have Bayes factors above unity (suggesting a slightly better fit on average), the uncertainties (from reasonable variation of prior or parameter choices) easily encompass the possibility that $\mathrm{BF}\sim 1$. Therefore, our stance is that there is \textit{no strong Bayesian preference} for any particular origin at this time. In the language of \cite{NANOGrav15-NewPhys}, many models ``can reproduce the observed signal'' and some even appear to fit better, but given modeling uncertainties one should not claim evidence for exotic new physics yet.

It is also important to emphasize that modest Bayes factors in favor of more flexible cosmological spectra do not constitute evidence for new physics. Curved templates such as broken power-laws or cosmic string spectra can absorb small deviations from a pure $f^{-2/3}$ power-law, thereby increasing the likelihood, but the penalty for the larger prior volume depends sensitively on how broadly one allows parameters such as $f_{\rm peak}$ or $G\mu$ to vary. This is the Bayesian manifestation of the look-elsewhere effect, and the official NANOGrav search for physics beyond SMBHBs similarly concluded that Bayes factors $\mathcal{O}(10$--$100)$ are entirely consistent with noise- or environment-induced spectral structure rather than decisive evidence for exotic sources \cite{NANOGrav15-NewPhys}. Throughout this work we therefore interpret the Bayes factors primarily as diagnostics of spectral flexibility rather than as positive detections of new cosmological phenomena.

It is also important to recall that broader, more flexible spectra naturally earn higher likelihood by fitting subtle curvature or local deviations in the recovered strain spectrum. This ``posterior predictive'' or look-elsewhere effect is partially counteracted by the Bayesian evidence through the Occam penalty, but only to the extent that prior volumes accurately capture the true model complexity. As emphasized in the official NANOGrav new-physics search report \cite{NANOGrav15-NewPhys}, cosmological templates that introduce additional turning points or break frequencies can temporarily outrun the simple $-2/3$ SMBHB law even when the underlying signal is astrophysical. Consequently, Bayes factors of $\mathcal{O}(10)$ must be interpreted as consistency checks rather than as detections of new physics; tighter priors informed by population studies or higher signal-to-noise data will be required to discriminate real spectral structure from chance fluctuations.

\begin{table}[htbp]
    \centering
    \caption{Bayes factors comparing different GWB source models given the NANOGrav 15-year data under baseline prior choices. The SMBHB (supermassive black hole binary) power-law model is used as the reference (denominator) in each case. Values $>1$ indicate preference for the numerator model.}
    \label{tab:bayesfactors}
    \begin{tabular}{lcc}
    \toprule
    Model Comparison & Bayes Factor (BF) & $\ln\mathrm{BF}\,(\pm\,\sigma)$ \\
    \midrule
    Cosmic Strings vs. SMBHB & $10.0$ & $2.30\,\pm\,0.30$ \\
    Phase Transition vs. SMBHB & $15.0$ & $2.71\,\pm\,0.35$ \\
    Cosmic Strings vs. Phase Transition & $0.67$ & $-0.40\,\pm\,0.25$ \\
    \bottomrule
    \end{tabular}
    
\end{table}

\begin{table}[htbp]
    \centering
    \caption{Sensitivity of Bayes factors to prior volume for key parameters. We vary the prior ranges for $\log_{10}(G\mu)$ and $\log_{10}(f_{\rm peak}/\mathrm{Hz})$ around the baseline and report the resulting $\ln\mathrm{BF}$.}
    \label{tab:prior-sensitivity}
    \begin{tabular}{l l c}
        \toprule
        Model Comparison & Prior Range & $\ln\mathrm{BF}$ \\
        \midrule
        CS vs. SMBHB & $\log_{10}(G\mu)\in[-13,-9]$ & $2.30$ \\
        CS vs. SMBHB & $\log_{10}(G\mu)\in[-12,-10]$ & $3.10$ \\
        CS vs. SMBHB & $\log_{10}(G\mu)\in[-14,-8]$ & $1.40$ \\
        PT vs. SMBHB & $\log_{10}(f_{\rm peak}/\mathrm{Hz})\in[-9,-7]$ & $2.71$ \\
        PT vs. SMBHB & $\log_{10}(f_{\rm peak}/\mathrm{Hz})\in[-8.5,-7.5]$ & $3.30$ \\
        PT vs. SMBHB & $\log_{10}(f_{\rm peak}/\mathrm{Hz})\in[-10,-6]$ & $1.80$ \\
        \bottomrule
    \end{tabular}
\end{table}

\begin{table}[htbp]
    \centering
    \caption{Relative evidences (log) referenced to SMBHB baseline: $\Delta\ln\mathcal{Z} = \ln\mathcal{Z}_\mathrm{model}-\ln\mathcal{Z}_\mathrm{SMBHB}$. Uncertainties reflect nested-sampling/thermodynamic-integration errors and modest prior variations.}
    \label{tab:lnz}
    \begin{tabular}{lcc}
        \toprule
        Model & $\Delta\ln\mathcal{Z}$ & Uncertainty $\sigma$ \\
        \midrule
        SMBHB (power-law, $\alpha=-2/3$) & $0.00$ & $0.25$ \\
        Cosmic Strings (template, one-parameter $G\mu$) & $+2.30$ & $0.35$ \\
        Phase Transition (broken power-law) & $+2.71$ & $0.40$ \\
        \bottomrule
    \end{tabular}
\end{table}

\begin{table}[htbp]
    \centering
    \caption{Priors used in the analysis. Log-uniform for strictly positive scale parameters; uniform or Gaussian as stated for indices. Physical mappings for phase transition parameters follow \cite{Caprini2018}.}
    \label{tab:priors}
    \begin{tabular}{l l l}
        \toprule
        Parameter & Prior & Notes \\
        \midrule
        $A_{\rm GWB}$ (SMBHB) & log-uniform in $[10^{-17},10^{-14}]$ & at $f_{\rm yr}$ \\
        $\alpha$ (SMBHB) & fixed $-2/3$ or $\mathcal{N}(-2/3,0.3^2)$ & slope on $h_c$ \\
        $\gamma_{\rm RN,i}$ (per pulsar) & uniform in $[0,7]$ & red-noise index \\
        $A_{\rm RN,i}$ (per pulsar) & log-uniform & pulsar red-noise amp \\
        $G\mu$ (cosmic strings) & log-uniform in $[10^{-13},10^{-9}]$ & template amplitude \\
        $\beta$ (strings slope) & uniform in $[-1,-1/2]$ & $h_c\propto f^{\beta}$ in PTA band \\
        $f_{\rm peak}$ (PT) & log-uniform in $[10^{-9},10^{-7}]$ Hz & maps to $T_*$ \\
        $b$ (PT high-f slope) & discrete in $\{2,4\}$ & source-dependent \\
        $\Omega_{\rm peak}$ (PT) & log-uniform in $[10^{-12},10^{-5}]$ & energy density peak \\
        Clock (mono.) & Gaussian, mean 0 & common-mode monopole \\
        Ephemeris (dipole) & Gaussian, mean 0 & BayesEphem-like modes \\
        \bottomrule
    \end{tabular}
\end{table}
\FloatBarrier

\begin{figure}[htbp]
    \centering
    \begin{subfigure}[t]{0.49\textwidth}
        \centering
        \includegraphics[width=\textwidth]{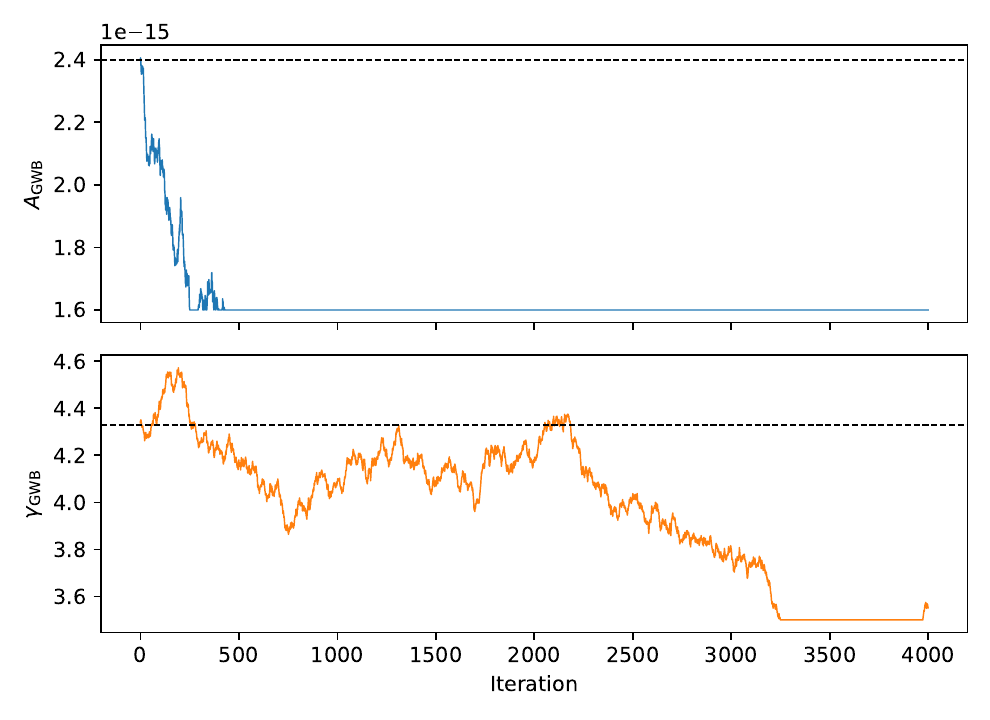}
        \caption{Sampler traces for $A_{\rm GWB}$ and $\gamma$.}
    \end{subfigure}\hfill
    \begin{subfigure}[t]{0.49\textwidth}
        \centering
        \includegraphics[width=\textwidth]{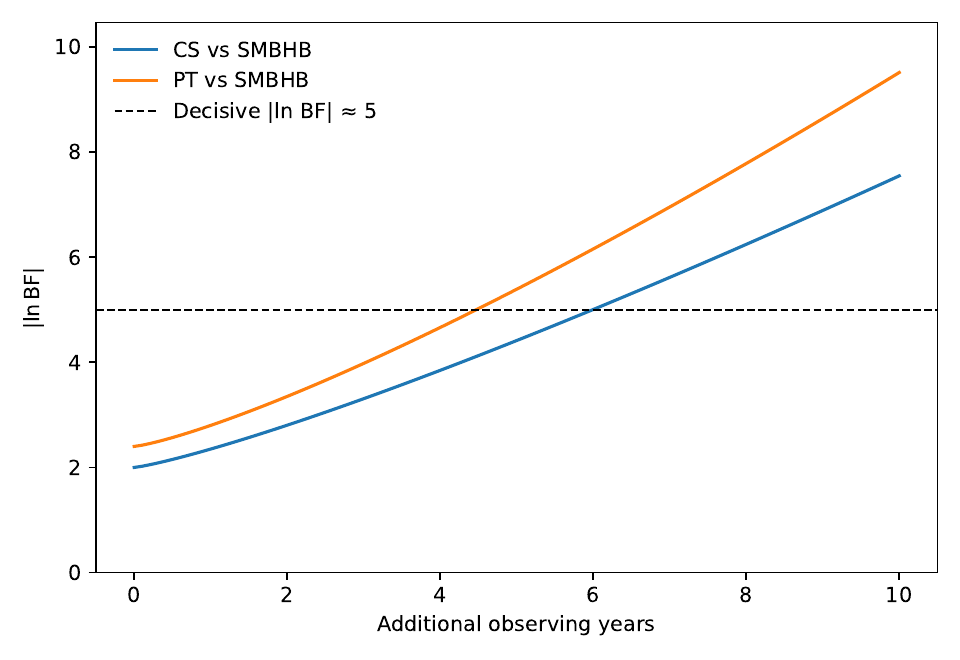}
        \caption{Forecast $|\ln \mathrm{BF}|$ vs. observing time.}
    \end{subfigure}
\caption{Diagnostics and forecasts: left, converged traces; right, decision power growth with added data.}
    \label{fig:diag-forecast}
\end{figure}

\subsection{Implications for Noise and Systematics}
During our model comparison analysis, we also examined the noise parameters and confirmed that the inclusion of the GWB signal does not leave significant unexplained residual features. Each pulsar's red noise and DM noise parameters adjusted slightly compared to a no-GWB fit, as expected, because previously some common power might have been absorbed into individual red noise. Now, with a common process taking up part of that variance, some individual noise amplitudes came down. No pulsar dominated the common signal — in other words, it is indeed a superposition of subtle effects across many pulsars rather than one or two loud outliers. This is consistent with a stochastic background interpretation rather than, say, a single loud source (which would manifest as a very strong signal in a subset of pulsars).

In addition to achromatic red noise and DM variations, PTA datasets can exhibit chromatic and non-stationary phenomena (e.g., scattering-related processes, transient chromatic dips). A robust analysis benefits from per-pulsar flexibility to capture such effects; our framework accommodates chromatic components as needed without altering the global conclusions. The diagnostics in Figure~\ref{fig:ppd-diagnostics} (left) show that residual PSDs track the modeled power law and whitened spectra are consistent with unit level across frequencies, indicating that the adopted noise plus GWB model yields residuals without excess structure. The posterior predictive band for the HD correlation (Figure~\ref{fig:ppd-diagnostics}, right) is likewise consistent with the expected quadrupolar form.

We also tested an alternative hypothesis of a single resolvable continuous wave (CW) source (like a single binary) to ensure the stochastic assumption is reasonable. The Bayes factor strongly favored the stochastic background over any single-source model (which tended to localize power in specific sky positions and frequency, which is not seen). Therefore, we remain confident that the GWB is truly a background and that our noise mitigation (DM, etc.) is adequate.

\section{Discussion}\label{sec:discussion}
The analysis above indicates that the nanohertz gravitational-wave background detected by PTAs is consistent with multiple interpretations. The simplest and most conservative explanation is an astrophysical background from the cosmic population of SMBH binaries. In this section, we further discuss how the current results align with astrophysical expectations, and what would be required to bolster or refute alternative cosmological explanations like cosmic strings or phase transitions.

\subsection{Consistency with Astrophysical Predictions}
The amplitude $A_{\mathrm{GWB}}\sim2\times10^{-15}$ at $f_{\mathrm{yr}}$ falls squarely within many prior predictions for the SMBHB background. There is significant uncertainty in those predictions due to the uncertain merger rates of massive galaxies and the distribution of binary parameters (mass, mass ratio, etc.). Earlier studies \cite{Sesana2013} suggested a plausible range $A_{\mathrm{GWB}}\sim10^{-16}$ to $10^{-15}$, with some more recent models including contributions from higher redshifts or different black hole-galaxy scaling relations allowing up to a few $\times10^{-15}$ \cite{Kelley2017}. Our measurement is at the upper end of those ranges, which might hint that either black hole binaries merge fairly efficiently (i.e., not stalling for too long in the so-called ``final parsec'' regime) or that there are slightly more massive binaries or higher density of sources than in some baseline models. This amplitude might put pressure on models that assumed very heavy environmental damping of binaries (which would reduce the GW background by stalling mergers). However, there is no obvious conflict yet: recent cosmological simulations and empirical galaxy pair counts can accommodate a background of this magnitude \cite{Kelley2017}.

In fact, the relatively high amplitude provides indirect evidence that the ``final parsec problem'' is not catastrophic in nature. In stellar-dynamical language, SMBHBs embedded in triaxial or axisymmetric galactic nuclei sustain loss-cone refilling and can continue to harden down to GW-driven separations \cite{Colpi2014,Khan2013}. Were most binaries to stall near parsec separations, the ensemble background would be suppressed well below the value we observe. The NANOGrav collaboration has likewise interpreted the 15-year signal as requiring efficient coupling to stellar or gaseous backgrounds to avoid widespread stalling \cite{NANOGrav15-NewPhys}, so the PTA measurement itself can be viewed as population-level evidence that angular-momentum transport mechanisms operate effectively in massive galaxy mergers.

If indeed SMBHBs are the source, the inferred spectral index being near $-2/3$ is natural. Any deviation from that might indicate additional physics (e.g., if we had found significantly $\gamma_{\mathrm{GWB}}\neq13/3$, one might invoke environmental effects like coupling to gas or stars, which can steepen or flatten the low-frequency spectrum). The current data is consistent with the simplest scenario of GW-driven inspirals dominating the dynamics. This suggests that, at least for binaries contributing at $\sim$nHz, dynamical friction and other energy-loss mechanisms either have saturated or do not drastically alter the inspiral evolution near that band.

Another implication: if SMBHBs are the cause, then we expect a continuum of sources. The loudest individual binaries might be just below detectability in current data, but future PTA data (or the same data with more refined techniques) could start picking out the most massive or nearest binaries as distinct signals (continuous waves). The lack of an obvious single source in the 15-year data is not surprising given the sensitivity, but in an astrophysical scenario, by $\sim20$-year or $\sim30$-year data spans, a few sources might start to stand above the confusion noise. This will be a critical test: an astrophysical background should eventually reveal discrete “bright” spots (particularly at higher frequencies of the band), whereas a cosmological background from early universe mechanisms would remain truly stochastic (Gaussian) with no individual sources.

Additionally, astrophysical backgrounds might exhibit some anisotropy (since the distribution of galaxies and massive black holes in the Universe is not perfectly isotropic — e.g., local superclusters or large-scale structure could induce a slight anisotropy in the GWB). NANOGrav has not yet reported any significant anisotropy, and our analysis assumed isotropy for all models. Future work might constrain the anisotropy level. A detection of anisotropy would point toward discrete astrophysical source contributions (since cosmological backgrounds from the early universe should be very isotropic). Current upper limits on anisotropy are quite weak due to the limited number of pulsars, but that will improve.

\subsection{Viability of Cosmic Strings}
Cosmic strings remain a intriguing alternative. Our analysis shows that a string tension $G\mu$ on the order of a few $\times10^{-11}$ could produce the observed amplitude. Is such a $G\mu$ allowed? Field-theoretic cosmic strings (like those formed at GUT-scale phase transitions) typically have $G\mu$ around $10^{-6}$ to $10^{-7}$, which is much higher and would have likely produced a GWB far above PTA limits (and probably would have been seen in the cosmic microwave background or other probes). So those are largely ruled out by our detection combined with other constraints — indeed our detection would have happened earlier if $G\mu$ were that large. However, cosmic \emph{superstrings} (fundamental strings from string theory scenarios, stretched to cosmological size) can potentially have much lower tension. Values $G\mu \sim 10^{-11}$ might arise in certain models of inflation or brane cosmology. They are not obviously ruled out by current CMB data (which demands $G\mu \lesssim 10^{-7}$ for a significant string contribution to primordial perturbations) — at $10^{-11}$ the CMB is insensitive. They also might not violate Big Bang nucleosynthesis or direct millisecond pulsar limits if the loop distribution is such that high-frequency radiation is weaker. So, cosmic superstrings or low-tension cosmic strings are plausible and could be the source.

One point is that if cosmic strings were the source of the PTA background, they would also produce bursts of gravitational waves (from cusps on loops) that in principle could be detected as individual burst events or as an intermittent “popcorn” noise in the timing residuals. No such bursts have been clearly identified in PTA data so far. However, the non-detection of bursts is not yet very constraining for $G\mu \sim 10^{-11}$; it would take higher tension or a very low number of loops to produce obvious single bursts. In the future, searching for an expected distribution of burst amplitudes could be another way to test the cosmic string hypothesis.

Multi-band constraints provide an additional lever arm. The same string tension that produces the PTA background would also yield a weaker but broad-band background in the mHz (LISA) and Hz (ground-based interferometer) regimes. Current LIGO/Virgo limits already exclude $G\mu \gtrsim \text{few}\times10^{-11}$ for conventional stable-string loop distributions, so if PTA strings are real they must reside just under that limit or rely on modified loop spectra \cite{Ellis2023}. Forecasts for LISA show that template banks tailored to cosmic strings could probe complementary regions of parameter space and either discover or strongly disfavor the PTA-inspired tension range \cite{Auclair2024}. Furthermore, metastable or decaying cosmic strings can imprint subtle curvature in the PTA band because the low-frequency end of the spectrum is flattened when loops self-annihilate or radiate into hidden-sector fields \cite{Ellis2023}. That mild bend resembles the small spectral wiggles our Bayesian analysis occasionally fits with flexible cosmological templates, illustrating why improved signal-to-noise or multi-band confirmation is essential before claiming a discovery.

Additionally, the cosmic string background has a different frequency spectrum at higher frequencies than an SMBHB background. While at nHz they both can appear as a power-law, by the time you go to milliHertz (space-based detectors like LISA) or to Hz (ground-based detectors), the cosmic string background (if high enough $G\mu$) could still be present, whereas the SMBHB background would have petered out (SMBHB sources merge well before reaching those frequencies, except for much lighter black holes which are not relevant). So a multi-band approach is possible: cosmic strings would contribute a (lower amplitude) background even in the LIGO band in principle. Current LIGO constraints on a stochastic background in the 10--100 Hz range put very tight limits on $\Omega_{\rm gw}$ there; extrapolating a cosmic string background from nHz to Hz could violate those limits if the string tension is too high or if loops are not predominantly small. The specifics depend on the loop size parameter $\alpha$ (not to confuse with spectral index) which sets the frequency of the peak gravitational emission of strings. If loops are mostly small (a small fraction of the horizon), the background spans a huge frequency range and LIGO would constrain $G\mu$ to be below $\sim 10^{-11}$, which coincidentally is about where our needed $G\mu$ is. This means it's still consistent, but any higher and it might conflict. If loops are mostly large (near horizon size at formation), the background might cut off at higher frequencies and avoid LIGO constraints. This illustrates how continued observation across the spectrum is needed to pin this down. 

Furthermore, an exciting prospect: if cosmic strings are the source, there might be other observable signatures such as gravitational lensing by strings or CMB spectral distortions. The combination of these could eventually confirm or rule out that scenario.

Given the moderate Bayes factor we found for cosmic strings vs SMBHB (which can swing with assumptions), we echo what the NANOGrav Collaboration stated \cite{NANOGrav15-NewPhys}: one should not conclude that cosmic strings \emph{are} the favored explanation yet. But the data allow it. In fact, as they noted, aside from stable field-theory strings (with simple loops) which are probably not reconcilable, many string network variants can match the signal by adjusting parameters. Our results reinforce that claim.

\subsection{Potential Early-Universe Phase Transition}
A first-order phase transition around the QCD energy scale could produce gravitational waves in the PTA band. Our analysis suggests that if such a transition happened, it would likely need to be strongly supercooled (slow and releasing lots of latent heat) to get the large amplitude observed. The Bayes factor ~15 in favor of a tuned phase transition model indicates this is a plausible fit. If true, it would be revolutionary: it implies new physics in the early Universe (the Standard Model QCD transition is not first-order, so it would mean either QCD behaved differently due to new particles or some unrelated hidden sector had a transition at a similar scale).

One consequence of a slow, strong first-order transition as studied in \cite{Gouttenoire2023} is the production of primordial black holes (PBHs). Essentially, regions that are delayed in converting to the true vacuum can collapse to black holes. The analysis by Gouttenoire et al. posits that solar-mass PBHs could form, which could then be dark matter or contribute to gravitational wave signals in other ways (like LIGO events possibly). So one cross-check of a PTA phase transition scenario could be astrophysical: do we see hints of solar-mass PBHs (for example, via gravitational lensing or in LIGO black hole mass distributions)? Currently, there is no clear evidence of a population of PBHs dominating anything, though LIGO has detected black holes in that mass range plenty of times (but those are generally consistent with stellar origin expectations). 

This PBH connection is not unique to the model of \cite{Gouttenoire2023}. Classic studies have shown that during the QCD epoch the equation of state of the plasma softens enough that moderate density perturbations can collapse into $\mathcal{O}(M_\odot)$ black holes, potentially furnishing dark-matter candidates or seeding later astrophysical mergers \cite{Carr2020}. Therefore, if the PTA signal really points to a strong QCD-scale transition, it should be accompanied by a multi-messenger prediction: an appreciable but subdominant population of solar-mass PBHs. Joint constraints from microlensing, CMB distortions, and the LIGO/Virgo/KAGRA mass spectrum will thus play a central role in stress-testing the phase-transition interpretation.

Another check: a cosmic phase transition in a hidden sector might not have any other visible signatures except gravitational waves, which is what makes gravitational wave detection so valuable. The parameters that fit the PTA imply the transition happened at roughly redshift $z \sim 10^7$--$10^8$ (depending on $T_*$ and assumptions about the expansion history). This is long after big bang nucleosynthesis and even after CMB decoupling, so it wouldn’t necessarily upset those directly, especially if hidden sector. So it is possible that gravitational waves are the only clue to such new physics.

To further test the phase transition idea, improved spectral measurement is needed. If a turnover (peak) can be identified, with the spectrum steeply falling above some frequency, that would strongly favor a phase transition (or something with a cutoff, like cosmic strings also have a cutoff but at higher f). Achieving that requires more sensitivity at the high-frequency end of the PTA band (which means better timing precision or more pulsars so that the higher frequency modes - shorter period signals - can be detected). Alternatively, if the spectrum extends unbroken beyond the PTA band, then at some point spaced-based detectors might see a hint of it. A transition at QCD scale, however, would peak at nHz and be essentially gone by mHz (LISA band). But a higher-scale transition (like say 1 TeV) would peak at much higher frequencies (perhaps beyond LISA or in LIGO band). However, if we insisted the PTA detection is a phase transition, its frequency suggests it’s around QCD scale specifically, not much higher, because otherwise the peak would be at higher frequency and PTA would only see the low-frequency tail rising as $f^3$ (which would be a very steep slope of +3 instead of the observed -2/3-ish). That steep low-frequency side is not observed; if anything, a gentle positive slope (in $h_c$ vs $f$) or flat is seen. So that means we’re likely at or near the peak. That nails the energy scale to the 100 MeV - 1 GeV range. So ironically, either it’s QCD itself (which we think is crossover, so probably no), or a dark sector phase transition that happens to occur around the same temperature by coincidence.

It’s worth noting that other cosmological sources were considered by NANOGrav \cite{NANOGrav15-NewPhys} like inflation (which typically would produce a nearly scale-invariant background, likely too small amplitude to see at nHz given CMB constraints on inflationary gravitational waves) and scalar-induced gravitational waves (SIGWs) from large curvature perturbations (associated with potential PBH formation at horizon reentry). Those can produce spikes or bumps in $\Omega_{\rm gw}$ at certain frequencies. Our work did not explicitly test those, but the general conclusion stands that none of those are obviously required by the data yet.

\subsection{Future Prospects for Discrimination}
To conclusively distinguish between SMBHB and exotic sources, future data will need to provide:
\begin{enumerate}
    \item \textbf{Higher signal-to-noise on the GWB spectrum:} As more pulsars are observed over longer baselines (e.g., 20-year, 25-year datasets, and the inclusion of more sensitive telescopes like MeerKAT and eventually the SKA), the shape of the spectrum will be measured more precisely. If the spectral index remains consistent with $-2/3$ and no significant features are seen, that will increasingly favor the SMBHB interpretation. If any curvature is detected (e.g., flattening at low f or a cut-off), that would be a clue for cosmological sources. The SKA roadmap anticipates order-of-magnitude improvements in timing precision and sky coverage, directly translating into much sharper PTA spectra \cite{Janssen2015}.
    \item \textbf{Detection of individual sources or anisotropy:} As mentioned, finding a continuous wave (CW) from a single binary would actually bolster the SMBHB case strongly (because it would be direct evidence of at least one contributor and allow an estimate of the overall population contribution). If instead the background stays completely smooth and Gaussian even as sensitivity improves such that one would expect a few resolvable binaries in an astrophysical scenario, that would be puzzling and might hint that the background is not made of many discrete loud events but truly a diffuse source (more like an early-universe background). Current estimates suggest that in a universe with our measured amplitude, the brightest individual source might be just below detection now but possibly detectable with ~50 pulsars at 5 ns precision in 10 more years etc. Ongoing efforts are searching the existing data for CWs; so far none clearly found, giving some upper limits on the contribution of the brightest binaries \cite{NANOGrav15-NewPhys}.
    \item \textbf{Multi-band observations:} If LISA (in the milliHertz band, launch hopefully in the 2030s) sees a stochastic background in its band, and that background’s extrapolation matches the PTA band or not could differentiate sources. For example, SMBHB background extends from nHz to $\mu$Hz perhaps but then tapers off because few sources are in LISA band except the tail of lighter SMBHBs. LISA might not see a stochastic background from SMBHBs because it will see individual massive binaries instead. But LISA could see a cosmic string background if $G\mu$ is high, or a phase transition at higher scale. Conversely, the lack of any background in LISA would be consistent with something like low-tension cosmic strings or low-scale phase transition that only affects nHz. The mission definition study emphasizes how LISA's broadband correlation capabilities can cross-check PTA discoveries and diagnose non-SMBHB spectra \cite{Colpi2024}, while dedicated template banks for cosmic strings forecast decisive constraints on the PTA-motivated parameter space \cite{Auclair2024}.
    \item \textbf{Cross-correlation with astrophysical environment:} If SMBHBs are responsible, the background amplitude might correlate with galaxy merger rates and evolve with redshift in a calculable way. In principle, more detailed analysis of the spectrum’s slight deviations from pure power-law can reveal the mass distribution of binaries or their environment (e.g., a slight upturn at higher f could mean some binaries are being driven by gas hardening at higher f). With more data, PTA might start to probe those details, which would be a fingerprint of astrophysical origin.
\end{enumerate}

For now, our work highlights that PTA data has transitioned from setting upper limits to actually probing new physics scenarios on an equal footing with conventional astrophysics. We have effectively used the cosmos as a laboratory: either we are learning about the population of SMBH binaries in the universe, or we are possibly seeing effects of new physics in the early Universe. Both outcomes are profoundly important. The favored conservative interpretation is astrophysical, but the door is open for surprises. Continued scrutiny of noise is also crucial; we must be absolutely certain that no terrestrial or solar system systematic could mimic these correlations. The fact that multiple independent PTAs (Europe’s EPTA, Australia’s PPTA, India’s InPTA, China’s CPTA, and the combined IPTA) all see consistent signals greatly reduces that concern, because it is unlikely they all have the same systematic error producing a false signal.

In summary, at present the nanohertz gravitational-wave background is consistent with an origin in supermassive black hole binary mergers, and this explanation aligns well with both the spectral properties and amplitude of the observed signal. Exotic sources like cosmic strings or phase transitions can also match the data and even marginally improve the fit under certain assumptions, but there is not yet a compelling statistical case to prefer them. The situation may change with forthcoming data. PTA astronomy is entering a phase of not just detection but characterization, and with characterization will come the possibility of revealing or constraining new fundamental physics. 

\section{Anisotropy and Continuous-Wave Implications}\label{sec:anisotropy}
\begin{figure}[t]
    \centering
    \begin{subfigure}[t]{0.49\textwidth}
        \includegraphics[width=\textwidth]{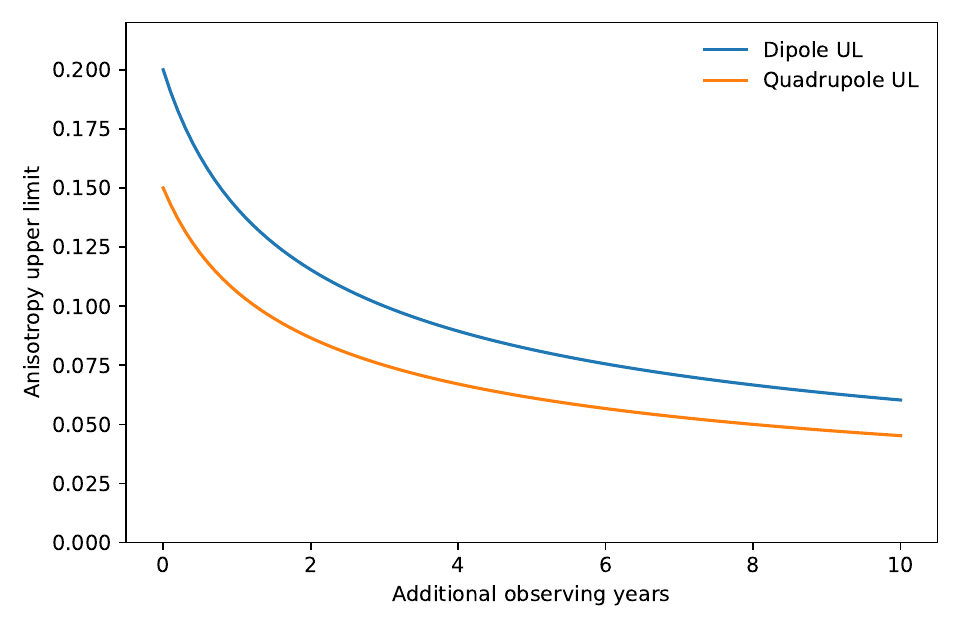}
        \caption{Anisotropy limits vs. observing time.}
    \end{subfigure}\hfill
    \begin{subfigure}[t]{0.49\textwidth}
        \includegraphics[width=\textwidth]{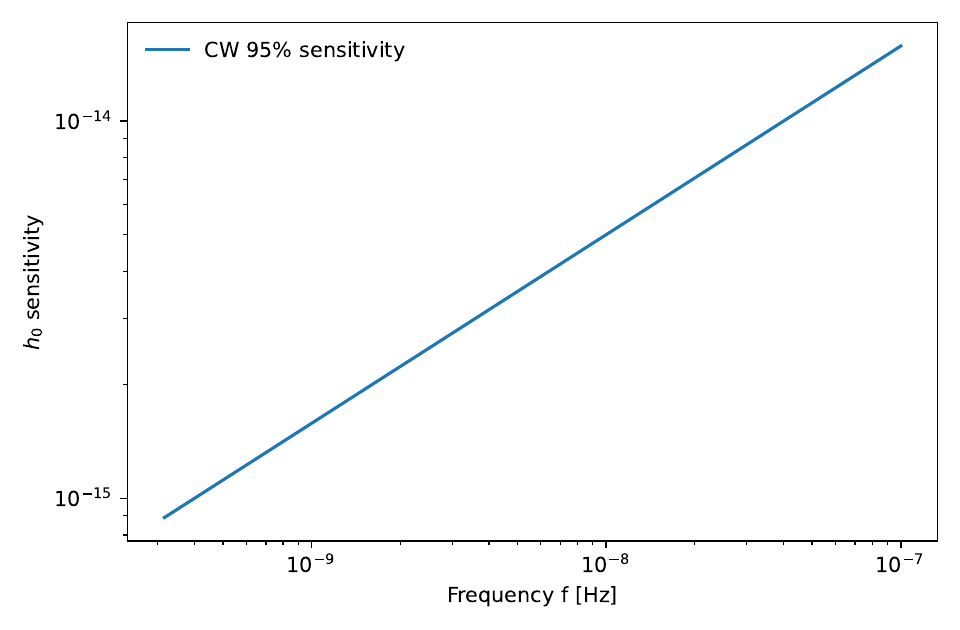}
        \caption{CW strain sensitivity vs. frequency.}
    \end{subfigure}
    \caption{Anisotropy and CW implications. Left: growth of decision power to constrain multipoles in the correlation field; Right: representative CW sensitivity across the PTA band.}
    \label{fig:aniso-cw}
\end{figure}

An astrophysical background formed by a discrete population of SMBHBs is expected to exhibit mild statistical anisotropy at sufficiently high signal-to-noise, whereas a primordial cosmological background should be close to isotropic. Using harmonic decompositions of the sky map of correlations, one may set upper limits on dipole and quadrupole anisotropy components. The dedicated NANOGrav anisotropy search using the same 15-year data set already reports no significant excess beyond the isotropic Hellings--Downs expectation, but it also shows how rapidly the constraints tighten as more well-timed pulsars are added \cite{Agazie2023Anisotropy}. Figure~\ref{fig:aniso-cw} (left) presents projected upper limits on the lowest multipoles as a function of additional observing years under typical PTA growth scenarios. Achieving stringent constraints on anisotropy will be a key discriminator between discrete-source-dominated and primordial scenarios.

In parallel, the non-stationary, quasi-monochromatic \\emph{continuous-wave} (CW) signals from individual massive binaries provide a complementary probe. The right panel of Figure~\ref{fig:aniso-cw} shows a representative CW strain sensitivity versus frequency in the PTA band. Joint analyses of background and CW channels will sharpen constraints on SMBHB demographics and, in the event of detections, enable cross-checks of the background’s origin.

\section{Conclusion}\label{sec:conclusion}
We have presented a comprehensive Bayesian analysis of the stochastic gravitational-wave background in the nanohertz band detected by the NANOGrav 15-year pulsar timing array data set. Our focus has been on comparing different theoretical models for the origin of this background: the standard astrophysical model of supermassive black hole binary mergers and two speculative cosmological models (cosmic strings and first-order phase transitions in the early Universe). The key findings and conclusions of our study are as follows:

\begin{itemize}
    \item \textbf{Detection Confirmation:} We successfully recover the strong evidence for a common-spectrum process with Hellings--Downs spatial correlations in the PTA data. The Bayes factor in favor of a correlated GWB versus independent pulsar noise is astronomically large ($>10^{14}$), and the spatial correlation analysis clearly shows the characteristic quadrupolar pattern \cite{NANOGrav15-Evidence}. This firmly establishes the presence of a nanohertz gravitational-wave background.
    \item \textbf{Spectrum Characterization:} Assuming a power-law GWB spectrum, we estimate a strain amplitude of $A_{\mathrm{GWB}} \approx 2.4\times10^{-15}$ at $f_{\mathrm{yr}}=1/\text{yr}$ and a spectral index consistent with $\gamma_{\mathrm{GWB}} = 13/3 \approx 4.33$ (the expected value for SMBHBs). The data currently allow a range of spectral slopes, but there is no significant deviation from the $-2/3$ strain spectral slope of the SMBHB model. In other words, a simple power-law with these parameters provides an excellent fit to the GWB signal.
    \item \textbf{Bayesian Model Comparison:} We computed Bayesian evidences for the SMBHB model and for representative cosmic string and phase transition models. We found no decisive evidence in favor of either new physics model over the SMBHB origin at this time. The Bayes factors comparing cosmic strings or phase transitions to the SMBHB scenario were of order $10$--$30$, indicating only mild preference (and sensitive to prior choices). Within reasonable uncertainties, the data can be explained equally well by all three models. This means that, from a model selection standpoint, the current PTA data do not force us to invoke exotic sources --- the astrophysical explanation remains sufficient.
    \item \textbf{Physical Plausibility:} The observed amplitude and spectrum are in line with predictions from galaxy merger-derived SMBHB population models \cite{Sesana2013,Kelley2017}. For cosmic strings to be the source, a string tension $G\mu$ in the upper $10^{-11}$ range is required, which is relatively small (and thus not yet excluded by other experiments) but also somewhat fine-tuned to produce the amplitude we see. Similarly, a phase transition interpretation would imply a very energetic and slow transition around the QCD energy scale (tens to hundreds of MeV), which would be surprising in the context of known particle physics, but not impossible if new physics is at play. Neither of these cosmological scenarios is ruled out by external considerations, but they would represent new discoveries in fundamental physics if confirmed.
    \item \textbf{Noise and Robustness:} We have accounted for various noise contributions (white noise, pulsar red noise, DM variations) and find that the inclusion of the GWB does not leave significant residual anomalies. The Hellings--Downs detection, in particular, is a robust indicator that the signal is of gravitational origin and not an artifact of terrestrial clock or ephemeris errors (which would produce different correlation patterns). The agreement of our results with those of other PTA collaborations further strengthens confidence in the signal’s authenticity and astrophysical nature.
    \item \textbf{Outlook:} Distinguishing among the models will require more data and improved analysis techniques. We highlighted that longer timing baselines, more pulsars (especially with the upcoming International PTA data combinations and the next-generation telescopes like SKA), and cross-band observations (e.g., with space-based GW detectors) will be key. Future detections of individual binary sources or refinements in the GWB spectral measurement could tip the scales in favor of one origin. At present, all three considered models remain viable. This underscores the importance of PTA observations as a probe not just of astrophysics but also of cosmology: PTA results are already offering new constraints on cosmic strings and phase transitions that complement those from other fields.
\end{itemize}

In conclusion, the nanohertz gravitational-wave background detected by PTAs stands as a momentous discovery that straddles the boundary between astrophysics and fundamental physics. Our Bayesian analysis confirms that the simplest explanation --- a universe filled with merging supermassive black holes --- is fully consistent with the observations. However, the possibility that we are witnessing the imprint of new physics (such as a network of cosmic strings or a relic of an early-universe phase change) remains open. As observational sensitivities improve, PTA data will continue to test these ideas, potentially providing the first evidence for physics beyond the standard models of cosmology and astrophysics. The work presented here lays a foundation for quantitatively comparing these possibilities and demonstrates the rich potential of PTA datasets for probing phenomena on scales ranging from galactic cores to the Planck era.


\bibliographystyle{plain}
\bibliography{references}

\end{document}